\documentclass[bibyear]{aa}

\usepackage{graphicx}

\usepackage{txfonts}

\usepackage{natbib}
\bibpunct{(}{)}{;}{a}{}{,}
\usepackage{booktabs}
\usepackage{multicol}
\usepackage{graphicx}
\usepackage{fancyhdr}

\usepackage{amsmath}	
\usepackage{amssymb}	
\usepackage[inter-unit-product=\cdot]{siunitx}
\usepackage{enumerate}
\usepackage{multirow}

\usepackage{mathtools}

\usepackage{longtable}
\usepackage{tabularx}
\usepackage{xcolor} 
\usepackage{ulem} 
\usepackage{comment}
\usepackage{cuted}

\usepackage{soul}

\usepackage{placeins}
\usepackage{stfloats}
\usepackage{float}

\usepackage{hyperref}
\hypersetup{colorlinks=true,urlcolor=black, citecolor=blue, linkcolor=black}

\makeatletter
\renewcommand*\aa@pageof{, page \thepage{} of \pageref*{LastPage}}
\makeatother

\newcommand{\orcidicon}[1]{\href{https://orcid.org/#1}{\includegraphics[width=11pt]{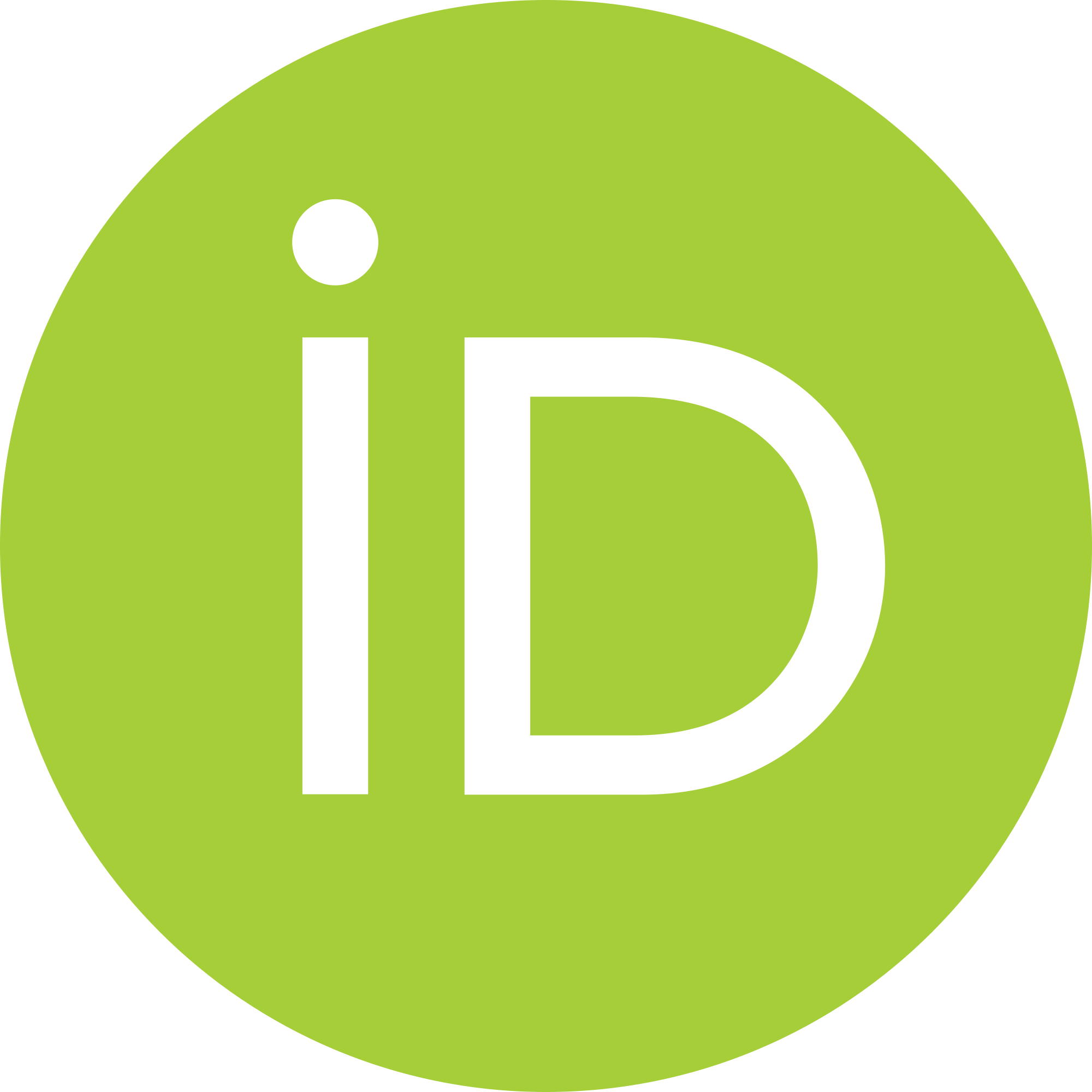}}}
\newcommand{\orcid}[1]{\href{https://orcid.org/#1}{\protect\orcidicon{#1}}}

\definecolor{steelblue}{rgb}{0.274 0.510 0.706}

\begin{document}

   \title{Metal-poor single Wolf-Rayet stars:\\ The interplay of optically thick winds and rotation}
    \titlerunning{}

   \author{
    Lumen Boco\inst{1}
    \orcid{0000-0003-3127-922X} \thanks{\href{mailto:lumen.boco@uni-heidelberg.de}{lumen.boco@uni-heidelberg.de}}, Michela Mapelli\inst{1,2,3}\orcid{0000-0001-8799-2548}\thanks{\href{mailto:mapelli@uni-heidelberg.de}{mapelli@uni-heidelberg.de}}, Andreas A. C. Sander\inst{4,2},  
    Sofia Mesini\inst{3,1}, Varsha Ramachandran\inst{4}, Stefano Torniamenti\inst{5,1}, Erika Korb\inst{3,1,6}, Boyuan Liu\inst{1}\orcid{0000-0002-4966-7450}, Gautham N. Sabhahit\inst{7}, Jorick S. Vink\inst{7}
    }
    \authorrunning{L. Boco et al.}
    \institute{
    $^{1}$Universit\"at Heidelberg, Zentrum f\"ur Astronomie (ZAH), Institut f\"ur Theoretische Astrophysik, Albert Ueberle Str. 2, 69120, Heidelberg, Germany\\
    $^2$Universit\"at Heidelberg, Interdiszipli\"ares Zentrum f\"ur Wissenschaftliches Rechnen, D-69120 Heidelberg, Germany\\
      $^3$Dipartimento di Fisica e Astronomia Galileo Galilei, Università di Padova, Vicolo dell’Osservatorio 3, I–35122 Padova, Italy
    $^{4}$Universit\"at Heidelberg, Zentrum f\"ur Astronomie (ZAH), ARI, Monchh\"ofstr. 12--14, 69120, Heidelberg, Germany\\
  $^5$Max-Planck-Institut f{\"u}r Astronomie, K{\"o}nigstuhl 17, 69117, Heidelberg, Germany\\
   $^6$INFN - Padova, Via Marzolo 8, I–35131 Padova, Italy \\
   $^7$Armagh Observatory and Planetarium, College Hill, Armagh BT61 9DG, Northern Ireland, UK\\}

   \date{Received XXXX; accepted YYYY}

\abstract{The Small Magellanic Cloud (SMC) hosts 12 known Wolf-Rayet (WR) stars, seven of which are apparently single. Their formation is a challenge for current stellar evolution models because line-driven winds are generally assumed to be quenched at a metallicity of $Z\leq{}0.004$. Here, we present a set of \textsc{mesa} models of single stars with zero-age main sequence masses of $20-80$~M$_\odot$ considering different initial rotation speeds ($\Omega=0-0.7\,{}\Omega_c$), metallicities ($Z=0.002-0.0045$), and wind mass-loss models (optically thin and thick winds). We show that if we account for optically thick winds, fast rotating ($\Omega\sim 0.6\,{}\Omega_c$) single metal-poor O-type stars (with $M\gtrsim{}20$ M$_\odot$) shed their envelope and become WR stars even at the low metallicity of the SMC. The luminosity, effective temperature, evolutionary timescale, surface abundance, and rotational velocity of our simulated WR stars are compatible to the WRs observed in the SMC. We speculate that this scenario can also alleviate the excess of giant stars across the Humphreys-Davidson limit. Our results have key implications for black hole masses, (pair instability) supernova explosions, and other observable signatures.} 
  

   \maketitle

\defcitealias{Sabhahit2023}{S23}
\defcitealias{Vink2001}{V01}

\section{Introduction}\label{sec:intro}
With a metallicity of $Z\sim{0.0025}$, the Small Magellanic Cloud (SMC) is a perfect laboratory to study the evolution of young, massive, and relatively metal-poor stars in the backyard of the Milky Way \citep{Vink2023xsu}. The SMC hosts 12 known Wolf-Rayet (WR) stars , i.e., hot ($T_{\rm eff}>40$~kK) massive stars whose spectrum is dominated by broad emission lines. At low metallicity, the formation of WRs is usually attributed to envelope stripping by a companion star \citep{Vanbeveren1997} because stellar winds are deemed to be too inefficient to (almost) completely remove the H-rich envelope \citep{Shenar2020}. However, observational searches have failed to identify a companion for seven out of the 12 known WRs in the SMC \citep{Westerlund1964, Smith1968, Sanduleak1968, Sanduleak1969, Breysacher1978, Azzopardi1979, Vanbeveren1980, Moffat1982, Moffat1985, Morgan1991, Bartzakos2001, Massey2001, Massey2003, Foellmi2003, Foellmi2004, Hainich2015, Shenar2016, Shenar2018, Neugent2018, Schootemeijer2024}. In particular, with a 95\% confidence level, \citet{Schootemeijer2024} rule out the presence of companions more massive than 5 M$_\odot$ and orbital periods shorter than one year for all seven WRs. This result implies that massive single stars at SMC metallicity lose the majority of their hydrogen-rich envelope without requiring a binary companion.

Stellar-structure codes are currently not able to reproduce the observed properties of single WRs in the SMC \citep{Hainich2015,Schootemeijer2024}. Without invoking binary interactions, reproducing the population of WRs is a difficult task, even at solar metallicity, as the predicted Hertzsprung-Russell (HR) diagram positions are commonly off with respect to their effective temperatures and luminosities. A temperature discrepancy arises as evolutionary models tend to produce WRs that are too hot ($T_{\rm eff}>100$ kK) compared to the observed ones ($40<T_{\rm eff}/{\rm kK}<100$). This is also referred to as the WR radius problem \citep[e.g.,][]{Hillier1987,Langer1988,Hamann2006,Grassitelli2018}. Detailed atmosphere models taking into account the effect of the (hot) iron opacity bump on the wind dynamics \citep{Graefener2005,Sander2020a,Sander2023} could demonstrate that for stars close to the Eddington limit, strong winds with high mass-loss rates can be launched, which effectively cloak the hydrostatic layers of the star. In these cases, the effective temperature referring to the hydrostatic layers is significantly higher than the observed effective temperature, as the wind can remain optically thick out to many stellar radii. In regimes where the iron bump is not able to launch a wind \citep{Moens2022,Sander2023}, the proximity of the star to the Eddington limit could still lead to a considerable inflation of the envelope \citep[e.g.,][]{Grafener2011}, which also effectively makes the star look cooler. Hence, the temperature or radius discrepancy can -- at least partially -- be attributed to a problem of the structure models in predicting accurate effective temperatures for stars with optically thick winds.

The luminosity discrepancy, instead, comes from the fact that theoretical models, even at solar metallicity, are able to reproduce only the most luminous WRs ($\log{L/\rm L_\odot}>5.2$), whereas they struggle with the low-luminosity ones. This happens because they are able to peel off the envelope only for the most massive stars $M>40\,\rm M_\odot$, which have stronger winds, and not for initial masses in the range $20\,{\rm M_\odot}<M<40\,\rm M_\odot$. Some codes, such as \textsc{FRANEC} \citep{Chieffi2013}, succeed in peeling off the H-rich envelope for all the stars with initial masses of $M>20\,\rm M_\odot$ thanks to rotation, but only at Galactic metallicity ($Z\sim{0.009-0.015}$). Other codes struggle even at Galactic metallicity \citep[see][]{Sander2019,Hamann2019}. With significantly weaker winds at SMC metallicity, reproducing observed WRs with single stellar evolution seems almost impossible. 

Another aspect of the same problem is that current evolutionary models overpredict the number of over-luminous supergiants falling inside the Humphreys-Davidson (HD) limit \citep{Humphreys1979}. \citet{Gilkis2021} demonstrated that current stellar evolution models overproduce very luminous supergiant stars, even when varying parameters such as rotation, semi-convection, and overshooting. Such an overproduction of luminous supergiant stars in theoretical models might be due to the mass-loss prescription adopted for massive stars. In particular, if stellar winds are too weak, the star is not able to lose its envelope, and it inflates even during the core hydrogen burning phase and remains inside the HD limit until the end of its life \citep[see e.g.][and references therein]{Sabhahit2022}. 
 
Mass-loss plays a key role during the evolution of massive stars and can determine their fate. Hot massive stars lose mass through radiation-driven stellar winds. The CAK model provides the traditional description of radiation-driven winds \citep{Castor1975}. This model is able to match the fundamental properties of OB star winds \citep[e.g.][]{Friend1986, Pauldrach1986}, but, among other issues, it struggles at explaining the properties of the most massive stars, especially if they are close to their Eddington limit \citep{Grafener2008,Grafener2011,Vink2011}. While an increase of the mass-loss rate very close to the Eddington limit, i.e., $\Gamma_\text{e}\approx{1}$ (where $\Gamma_\text{e}\equiv{}L/L_e$ is the ratio between the stellar luminosity $L$ and the Eddington luminosity $L_e$), in principle is already expected from CAK, \citet{Vink2011} found a significant change of slope, a kink, for high $\Gamma_\text{e}$ with a very steep trend of  $\dot M\propto\Gamma_\text{e}^{4.77}$. Observationally, there is also a change of spectral appearance from Of to WNh-type, even on the main sequence \citep[e.g.,][]{deKoter1997,Crowther2011}, meaning that the stars are eventually classified as Wolf-Rayet stars due to their emission-line spectra independent of their evolutionary status. \citet{Bestenlehner2014} observationally studied this transition regime, confirming the findings from \citet{Vink2011} based on Monte Carlo wind models.

As we discuss further in Sect.\,\ref{sec:wind-mass-loss}, the change in the $\dot{M}(\Gamma_\text{e})$ behavior can be related to a switch from an optically thin to an optically thick wind regime. \citet{Sabhahit2022} developed a formalism to account for this switch and applied it to reconcile the temperature evolution of very massive stars ($M > 100\,\rm M_\odot$) to observations in the Milky Way and the Large Magellanic Cloud (LMC). In particular, they showed that by switching to the optically thick regime, very massive stars avoid envelope inflation and spend their main sequence evolution in a narrow range of temperatures that are in agreement with observations. \citet[hereafter \citetalias{Sabhahit2023}]{Sabhahit2023} extended this formalism to lower metallicities (down to $Z_\odot/100$) to study the threshold of pair-instability supernovae.

In this work, we show that the optically thick wind model of \citetalias{Sabhahit2023} is able to explain the self-stripping of single WRs at SMC metallicities when combined with fast stellar rotation. To avoid the luminosity discrepancy, we applied the \citetalias{Sabhahit2023} formalism to stars with a lower mass ($M\geq 20\,\rm M_\odot$). The paper is organized as follows:  Section \ref{sec:methods} presents our methods, describing the code used for our simulations and quickly summarizing the \citetalias{Sabhahit2023} formalism for the optically thick wind regime. In Section \ref{sec:first_results}, we show our main results and their dependence on metallicity and rotation. Section \ref{sec:properties} discusses additional observable features of our simulated stars (e.g., surface abundances). In Section \ref{sec:discussion}, we discuss some possible modifications and implications of our model. Finally, Section \ref{sec:summary} presents a summary of our work.

\section{Methods} \label{sec:methods}

\subsection{General setup}
We used the Modules for Experiments in Stellar Astrophysics code (\textsc{mesa}; version r12115, \citealt{Paxton2011, Paxton2013, Paxton2015, Paxton2018, Paxton2019}) to evolve stars with different zero-age main sequence (ZAMS) masses $M_\text{ZAMS}/\rm M_\odot=20,$ 25, 30, 40, 50, 60, 70, and 80. We simulated six different metallicities that might be representative of different regions of the SMC $Z=0.002,$ 0.0025, 0.003, 0.0035, 0.004, 0.0045, and eight initial rotation speeds $\Omega/\Omega_c=0,$ 0.3, 0.45, 0.5, 0.55, 0.6, 0.65, 0.7, with critical velocity $\Omega_c^2=(1-\Gamma_\text{e})\,G\,{}M/R^3$, where $\Gamma_\text{e}\equiv{}L/L_e=\chi_e\,L/(4\pi\,G\,c\,M)$ is the electron scattering Eddington parameter ($L_e$ and $\chi_e$ being respectively the Eddington luminosity and the electron scattering opacity), $G$ is the gravity constant, $M$ the stellar mass, and $R$ the stellar radius. The rotational mixing efficiency is set by the ratio of the turbulent viscosity to the diffusion coefficient $f_c$ (the \verb|am_D_mix_factor| parameter of \textsc{mesa}) and the ratio between the actual molecular weight gradient and the value used for computing the mixing coefficients $f_\mu$ (\verb|am_gradmu_factor|), calibrated as in \citet{Heger2000}: $f_c=1/30$, $f_\mu=0.05$. 

We used the \citet{Grevesse1998} initial composition and opacity tables. For the equation of state (EOS), we used the default tables available in \textsc{mesa}, that are mainly based on the OPAL EOS tables \citep{Rogers1996}. Convective regions are defined by the Ledoux criterion, with mixing length parameter $\alpha_{\rm MLT}=1.5$ and semi-convective diffusion coefficient $\alpha_{\rm sc}=1$. 

We divided our simulations into two steps. The first roughly corresponds to the main sequence and lasts until the central hydrogen fraction drops below $10^{-2}$, and the second represents a more advanced phase and lasts until the central temperature reaches the value $\log (T_c/\rm K)=9.55$. With this choice, we ran all of our stellar models at least until the onset of O-burning. During the first phase, we adopt the default nuclear reaction network in \textsc{mesa} (\verb|basic| net), whereas in the second phase we use the \verb|co_burn| net (parameter \verb|auto_extend_net = true|), which includes $^{28}$Si, in order to have a more complete and extended network for C- and O-burning and $\alpha$-chains with respect to \citetalias{Sabhahit2023}.

As in \citetalias{Sabhahit2023}, in the first evolutionary step, the \textsc{MLT++} prescription is switched off, while in the second step it is switched on. MLT++ is a numerical patch to standard MLT used in \textsc{mesa} to artificially reduce the superadiabatic gradient in radiation-dominated, convective envelopes. Following \citetalias{Sabhahit2023}, the diffusion coefficient in the overshooting region is described by the exponential criterion with overshooting coefficient $f_{\rm ov}=0.03$ in the first step. The choice of $f_{\rm ov}=0.03$ during the core hydrogen burning phase is motivated by several recent studies, which indicate that massive stars require increased convective boundary mixing than low-mass stars in order to match observations\footnote{From asteroseismology of the $\beta$-Cephei $\theta$ Ophiuchi, \citet{Briquet2007} derive a step overshooting parameter $\alpha_{\rm ov}\sim 0.44\pm 0.07$. \citet{Vink2010} derive a value of $\alpha_{\rm ov}\sim 0.335$, based on the absence of fast-rotating Galactic B supergiants below $22\,\rm kK$. By studying the distribution of rotational velocities, \citet{Brott2011} find values of $\alpha_{\rm ov}\sim 0.34\pm 0.1$ for a $16\,\rm M_\odot$ star in the SMC. Collecting literature sets from different atmosphere analyses of Galactic massive stars, \citet{Castro2014} conclude $\alpha_{\rm ov}\sim 0.34$ at $M\sim16\,\rm M_\odot$, and further suggest a mass dependence: weaker overshooting in lower-mass stars and stronger overshooting at higher masses. \citet{Schootemeijer2019} constrained internal mixing using RSG/BSG ratios in the SMC, finding $0.2 \leq \alpha_{\rm ov} \leq 0.3$. \citet{Higgins2019} report large overshooting values $\alpha_{\rm ov} \sim 0.5$ using the luminosity--mass plane to reproduce Galactic binaries. \citet{Costa2019a}, analyzing double-line eclipsing binaries, also find evidence for strong mixing, with $\alpha_{\rm ov}$ in the range $0.3$–$0.8$. We note that most of these studies employ a step overshooting model with parameter $\alpha_{\rm ov}$, which is approximately related to the exponential overshooting parameter $f_{\rm ov}$  by $f_{\rm ov}\sim\alpha_{\rm ov}/10$. All these evidences motivates us that the choice of $f_{\rm ov}=0.03$ is a conservative assumption for overshooting in massive stars.}. In Section \ref{sec:discussion} we discuss how variations of $f_{\rm ov}$ affect the stellar evolution models.

Mixing in later evolutionary stages, instead, is more difficult to constrain and remains highly uncertain. The He-burning core develops a much steeper composition gradient compared to the H-burning core. This strong chemical discontinuity above He core can restrict the growth of the convective region \citep{Langer1991}. For this reason we choose a smaller value of $f_{\rm ov}=0.01$ during the advanced phase, as done by \citetalias{Sabhahit2023}. We have verified that this choice produces only minor differences in the evolution of our stellar tracks. 

\subsection{Wind mass-loss}\label{sec:wind-mass-loss}
For the wind mass-loss, we closely followed the model implemented by \citetalias{Sabhahit2023}, of which we give a brief recap here. \citetalias{Sabhahit2023} implement a consistent procedure to find the switch point between optically thin and thick winds $\Gamma_\text{e,switch}$ as a function of stellar parameters, and adopt the steep $\Gamma_\text{e}$ scaling for optically thick winds found by \citet{Vink2011}. Specifically, \citetalias{Sabhahit2023} follow the formalism by \cite{Vink2012}, who found a model-independent way to characterize the transition from optically thin winds of O-type stars to optically thick winds of WNh stars, irrespective of the assumptions on the wind clumping factor. They derived a relation between the wind efficiency parameter, $\eta$, and the optical depth parameter, $\tau$:
\begin{equation}
\eta\equiv\frac{M\,\varv_\infty}{L/c}=f\,\tau.
\end{equation}
The switch between optically thin and thick winds occurs when $\tau=1$ and corresponds to a wind efficiency parameter $\eta_{\rm switch}=f\simeq 0.6$. 

\citet{Vink2012} have prescribed a way to derive the transition mass-loss rate, $\dot M_{\textrm{switch}}$, by combining this equation with empirical observations of luminosities and terminal velocities of slash (i.e., Of/WNh) stars. If applied to slash stars in the Arches cluster, at Galactic metallicity, the derived $\dot M_{\textrm{switch}}$ agrees well with the empirically determined one assuming a clumping factor of $D=10$ \citep{Martins2008} and with the optically thin mass-loss recipe from \citet{Vink2001}, confirming that such a model for optically thin winds correctly predicts the mass-loss rates at the transition. 

\citetalias{Sabhahit2023} employed a set of stellar atmosphere models computed with the dynamically consistent branch of the Potsdam Wolf–Rayet (PoWR) code \citep{Sander2017,Sander2020a} to study the dependence of $f$ on various stellar parameters. They find that $f$ mainly depends on the ratio $\varv_\text{esc}/\varv_\infty$, where $\varv_\text{esc}\equiv(2\,G\,M/R)^{1/2}$, and $\varv_\infty\equiv a\,[2\,G\,M\,(1-\Gamma_\text{e})/R]^{1/2}\,(Z/Z_\odot)^{0.2}$ \citep{Lamers1995}, with $a=2.6$ above the bi-stability jump ($T_{\rm eff}>25\,$kK) and $a=1.3$ below it ($T_{\rm eff}<25\,$kK). Specifically, \citetalias{Sabhahit2023} found an equation for the wind efficiency parameter at the switch:
\begin{equation}
\eta_{\rm switch}=f=0.75\,{}\left(1+\frac{\varv_\text{esc}^2}{\varv_\infty^2}\right)^{-1}.
\label{eq:eta_s}
\end{equation}
This equation was initially derived for very massive stars, $M>50\,\rm M_\odot$, and is in agreement with the $f$ values reported by \cite{Vink2012} for the stratification-wind parameter $\beta=1.5$ (see their Table 2), which is an appropriate value for stars of this mass. Since we are interested in applying the \citetalias{Sabhahit2023} formalism to stars with $M>20\,\rm M_\odot$, we checked what happens if we rescale the value $0.75$ in Eq.\,\eqref{eq:eta_s}\,{}{} to be in agreement with the values of $f$ reported by \cite{Vink2012} for $\beta=1$ at $\varv_\infty/\varv_\text{esc}=1.5,\,2.5$. We found negligible differences, and thus we use $\eta_{\rm switch}$ in our work.

By means of Eq.\,\eqref{eq:eta_s}, we can calculate the value of the Eddington factor $\Gamma_\text{e,switch}$ at which the transition from optically thin to optically thick winds takes place $\eta(\Gamma_\text{e,switch})=\eta_{\rm switch}(\Gamma_\text{e,switch})$, where the mass-loss at the transition is computed via the optically thin prescription by \citet{Vink2001} $\dot M=\dot M_\text{V01}(L,M, T, \varv_\infty, Z)$, $M$ is evaluated from the mass-luminosity relations under homogeneity assumption, and the radius is computed from the Stefan-Boltzmann law. The value of $\Gamma_\text{e,switch}$ is therefore determined by the functions $\eta$ and $\eta_{\rm switch}$. An increase in $\eta$ or a decrease in $\eta_{\rm switch}$ results in a lower $\Gamma_\text{e,switch}$ and vice versa.

As in \citetalias{Sabhahit2023}, in the case of rotation, we multiplied the wind mass-loss rate by a factor \citep[][]{Maeder2000}:
\begin{equation}
\frac{\dot M(\Omega)}{\dot M(\Omega=0)}=\frac{(1-\Gamma)^{\frac{1}{\alpha}-1}}{\left(1-\Gamma-\frac{4}{9}\,\left(\frac{\varv}{\varv_\text{crit}}\right)^2\right)^{\frac{1}{\alpha}-1}}
\label{eq:boost},
\end{equation}
where $\Gamma\equiv \chi\,L/(4\pi\,G\,c\,M)$ is the total Eddington parameter ($\chi$ being the total opacity), $\alpha=0.52$, and $\varv_\text{crit}=[2\,G\,M/(3\,R_\text{pb})]^{1/2}$, where $R_\text{pb}$ is the polar radius at breakup, which we assumed does not change with rotation.

As in \citetalias{Sabhahit2023}, we used the following recipes for wind mass-loss:
\begin{itemize}
\item For temperatures in the range $4000\,$K$<T_{\rm eff}<10^5\,$K, in the optically thin regime ($\eta<\eta_{\rm switch}$) the standard \citet{Vink2001} prescription is used, while in the optically thick regime ($\eta>\eta_{\rm switch}$), the \cite{Vink2011} prescription is used.
\item For temperatures $T_{\rm eff}<4000\,$K, we used the red supergiants prescription by \citet{deJager1988}.
\item For temperatures $T_{\rm eff}>10^5\,$K, we used the maximum between the WR wind prescription by \citet{Sander2020} and \citet{Vink2017}. We call these winds WR-type winds in the rest of the paper.
\end{itemize}
Throughout the paper, we compare the \citetalias{Sabhahit2023} wind model with a model where only optically thin winds with the recipe of \citet{Vink2001} are considered. We call this case \citetalias{Vink2001}.

\section{Results}
\subsection{Self-stripping of WRs at SMC metallicities}\label{sec:first_results}
\begin{figure*}[t]
    \centering
    \includegraphics[width=0.85\linewidth]{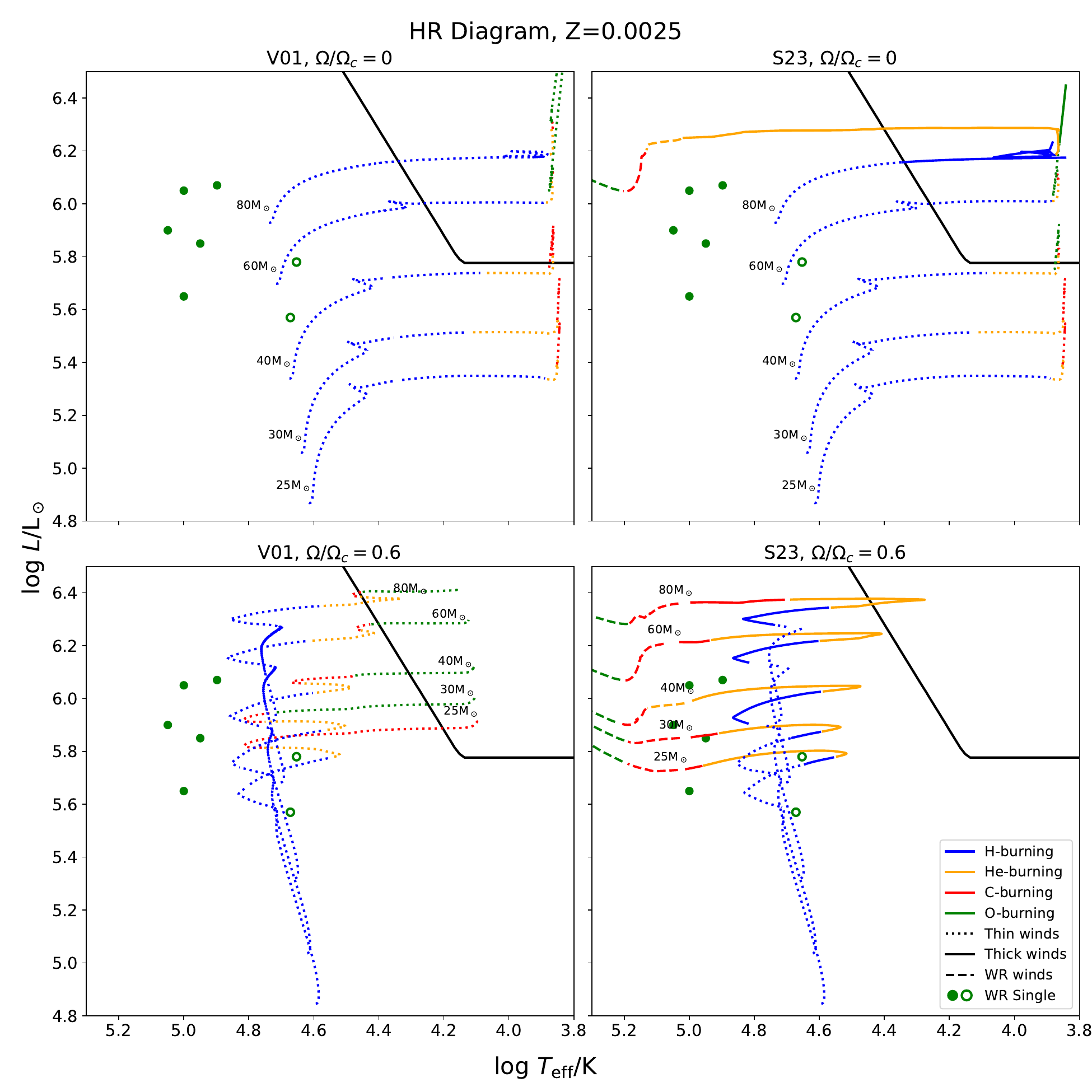}
    \caption{Stellar tracks on the HR diagram for initial masses $M=25,\,30,\, 40,\, 60$,\, and $80$~M$_\odot$ at metallicity $Z=0.0025$. In the left-hand panels, the \citetalias{Vink2001} model with only optically thin winds is implemented. In the right-hand panels, the \citetalias{Sabhahit2023} model is enforced, with the possibility to activate optically thick winds. The upper panels show the case with no rotation, while the lower panels the $\Omega/\Omega_c=0.6$ case. Different line styles represent different wind regimes: dotted for optically thin winds, solid for optically thick winds, dashed for WR-type winds. Colors represent different burning stages, defined as the element whose burning generates most of the energy of the star. Blue is for H-burning through CNO, orange for He-burning through triple $\alpha$, red for carbon burning, and green for oxygen burning. Green circles are observations of single WRs \citep{Hainich2015}, filled for the hottest, open for the coldest. The lower-right plot shows that the combination of optically thick winds and high initial rotation is able to self-strip stars even at SMC metallicity. Stars in the lower-right plot (\citetalias{Sabhahit2023}, $\Omega/\Omega_c=0.6$) avoid the HD limit and transition from main-sequence stars to WRs.}
    \label{fig:new_winds}
\end{figure*}
Figure \ref{fig:new_winds} shows our stellar tracks at metallicity $Z=0.0025$, which corresponds to the SMC baseline metallicity computed from the element abundances reported in  \cite{Vink2023xsu}. This figure compares stellar tracks in the \citetalias{Vink2001} case, where only optically thin winds are used (left), with the \citetalias{Sabhahit2023} case, where optically thick winds can be activated (right) if the condition $\Gamma_\text{e}>\Gamma_\text{e,switch}$ is met. We also compare models without and with rotation. For the cases with rotation, we assume initial rotation $\Omega/\Omega_\text{c}=0.6$, corresponding to an initial rotational velocity of $\varv_{\rm rot, i}\simeq 450$~km/s. 

Figure \ref{fig:new_winds} shows that optically thin winds at the metallicity of the SMC (top left) are not able to peel off the stellar envelope and the stars evolve to the cool side of the HR diagram after the main sequence, entering and remaining inside the HD limit if they are massive enough ($M\gtrsim 40\,\rm M_\odot$). 

A rotation as high as $\varv_{\rm rot, i}\simeq 450$ km/s (bottom left) changes the main sequence evolution of the star, which enters the chemically homogeneous evolution regime, becoming more luminous and more compact, but it is not enough to peel off the envelope. Even at this high initial rotation speed, if only thin winds are allowed, all stars end up being cool supergiants, entering the HD limit at the end of their life. 

Contrariwise, in the \citetalias{Sabhahit2023} model, some stars are able to switch to optically thick winds, peel off their hydrogen-rich envelope, move to the hot side of the HR diagram, and avoid the HD limit \citep{Romagnolo2024}. In the case with no rotation, this happens only for the most massive stars ($M\simeq 80\,\rm M_\odot$), as already shown by \citetalias{Sabhahit2023}. However, for rotating stars with $\Omega/\Omega_c=0.6$, this behavior extends to much lower initial masses ($M\simeq 25\,\rm M_\odot$). 

This happens because of several concurring factors. The most straightforward one is the boost factor multiplying the wind for rotating stars (Eq.\,\ref{eq:boost}). This increases the strength of the wind and the wind efficiency parameter $\eta$ and, consequently, reduces $\Gamma_\text{e,switch}$, as also shown in Fig. 7 by \citetalias{Sabhahit2023}.

However, the most important factors are related to the evolution of the star during the main sequence. Fast rotating stars, as the ones shown in Figure \ref{fig:new_winds}, bottom panels, evolve by chemically homogeneous evolution and become more luminous and more compact than non-rotating stars. The increase in luminosity has a twofold effect: (i) $\Gamma_\text{e}$ increases, (ii) the wind strength $\dot M$ and wind efficiency parameter $\eta$ increase, reducing $\Gamma_\text{e,switch}$. Furthermore, the smaller radius of rotating stars contributes to the increase of $\varv_\infty$ and consequently to a further increase of $\eta$, which reduces $\Gamma_\text{e,switch}$. All these effects concur in facilitating the activation of optically thick winds, even for less massive stars. If optically thick winds are activated before the onset of core He-burning, the star is able to peel off its hydrogen-rich envelope and becomes hot and compact, reaching temperatures $>10^5$~K and avoiding the HD limit. 

Figure \ref{fig:new_winds} demonstrates that the combination of optically thick winds and rotation might be the key to reproduce single WR observations in the SMC, especially the five hottest single WR stars highlighted by the solid green circles in the Figure\footnote{The positions on the HR diagram of the two coldest WRs (open circles) can be reproduced by all the models. However, non-rotating models cannot reproduce their evolved evolutionary stage, with severe surface hydrogen depletion. Therefore, only the two models with chemical homogeneous evolution can reproduce the coldest WRs \citep[see][]{Martins2009}.}. In our models, we overcome the claimed luminosity discrepancy as stars with initial mass as low as $25\,\rm M_\odot$ can self-strip their envelope. We can reproduce the empirical HR diagram location of the single WRs with $5.6<\log(L/\rm L_\odot)<6$ with rapidly rotating stars with initial masses in the range $25<M/\rm M_\odot<40$. 

The ability of stars in this mass range to lose their envelope also helps to address another issue: the observational lack of massive WR progenitors in the SMC. Current evolutionary models predict that  luminous single WRs at low $Z$ originate from very massive O-type stars ($>40\,\rm M_\odot$). However, observations \citep{Ramachandran2019, Schootemeijer2021} reveal a scarcity of such massive O-type stars in the SMC. Furthermore, even some of the seemingly luminous O-type stars in the SMC are actually stripped and have lower masses \citep{Pauli2022}. Given that the helium burning WR phase constitutes only about $\sim 10\%$ of a star's lifetime, we would expect to observe hundreds of massive O-type stars to account for the current WR population. This discrepancy suggests that WRs in the SMC likely evolved from less massive O-type stars, consistent with our findings that enhanced envelope stripping via thick winds can enable such evolutionary pathways.
 
\begin{figure}
    \centering
    \includegraphics[width=0.9\linewidth]{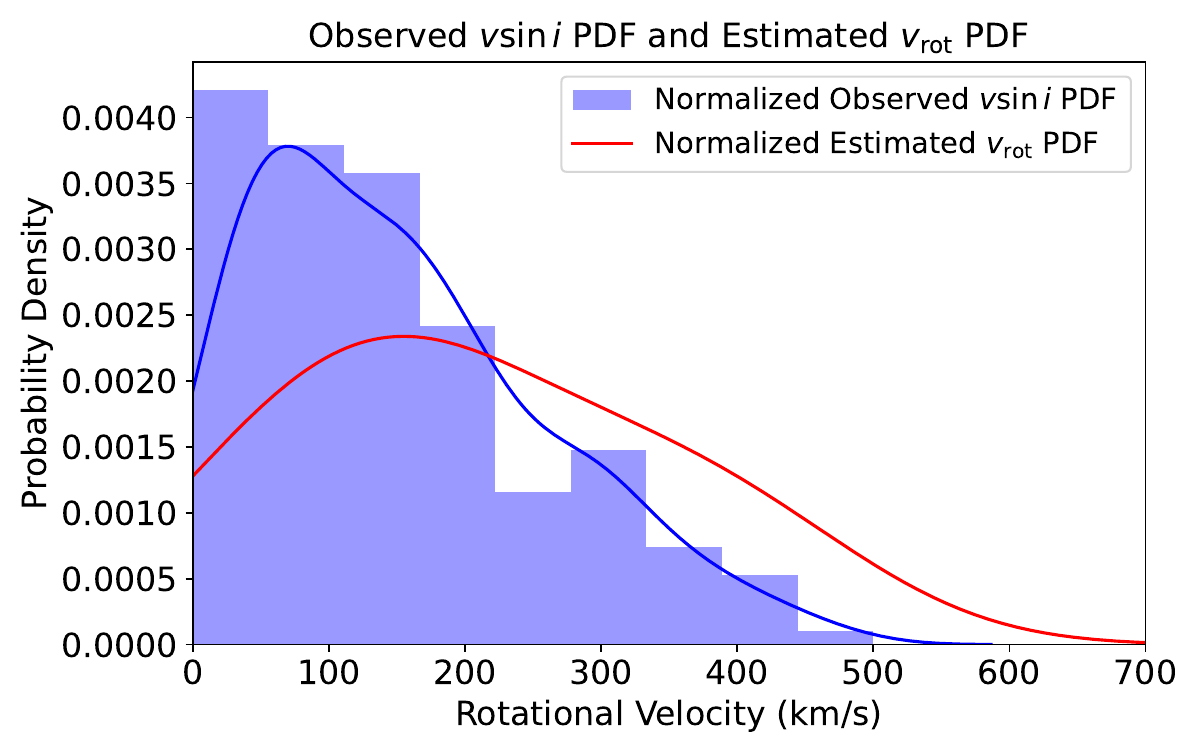}
    \caption{Velocity distribution of O-type stars in the SMC. The plot shows the normalized observed $\varv \sin i$ distribution (blue) and the reconstructed initial rotational velocity ($\varv_{\rm rot}$, red) for 171 O-type stars. The dataset is a compilation from \citet{Ramachandran2019}, \citet{Rickard2024}, and \citet{Bestenlehner2025}. About $10\%$ of the O-type stars feature $\varv_{\rm rot}>450$ km/s.}
    \label{fig:velocity_distribution}
\end{figure}

Rapidly rotating stars with initial velocity $\Omega/\Omega_c\gtrsim0.6$, corresponding to $\varv_{\rm rot}\gtrsim 450$ km/s, are not representative of the bulk of the O-type stars' rotational velocity distribution in the SMC.  We compiled $\varv \sin i$ values estimated for SMC O-type stars based on the recent literature from the BLOem survey \citep{Bestenlehner2025}, which covers the main bar of the SMC, combined with data from \citep{Ramachandran2019, Rickard2024}, that primarily cover the SMC wing and the core of the NGC346 star cluster. This combined dataset resulted in a total of 171 O-type stars, and the distribution of their  observed $\varv \sin i$ and true rotational velocities ($\varv_{\rm rot}$) are shown in Figure \ref{fig:velocity_distribution}. To determine the intrinsic rotational velocity distribution, we employed the Lucy-Richardson deconvolution method, an iterative technique that converges to a maximum likelihood solution, as demonstrated by \citet{Lucy1974}. Our analysis indicated that O-type stars with $\varv_{\rm rot}\gtrsim 450$~km/s approximately represent $\sim 10\%$ of the population in the SMC. Most of the observed O-type stars are no longer on the ZAMS, implying they may have already spun down. Therefore, even a small number of rapidly rotating O-type stars might be enough to explain the observed seven single WRs in the SMC. 

\subsubsection{The dependence on metallicity and rotation}\label{sec:dependence}
In this section, we aim to describe the complex dependence of the activation of thick winds on metallicity and initial rotation. We focus on stars with mass $M<40\,\rm M_\odot$, which are the most important to reproduce the observed WRs in the luminosity range $5.6<\log(L/L_\odot)<6$. In principle, one would expect weaker winds at a lower metallicity. While this remains generally true, the activation of the optically thick wind regime follows a more complex behavior. 
 
Stars at a lower metallicity are generally more compact and luminous than stars at a higher metallicity \citep{kippenhahn2013, Klencki2020, Gilkis2021}. This has two effects: On the one hand, a higher luminosity means a higher $\Gamma_\text{e}$. On the other hand, a higher luminosity and compactness increase the wind efficiency parameter, $\eta$,  reducing $\Gamma_\text{e, switch}$ and facilitating the activation of thick winds. This scenario is further complicated by the bi-stability jump \citep{Vink1999, Vink2000, Vink2001}, which enhances wind strength for stars cooler than 25~kK. 

\begin{figure}
    \centering
    \includegraphics[width=0.85\linewidth]{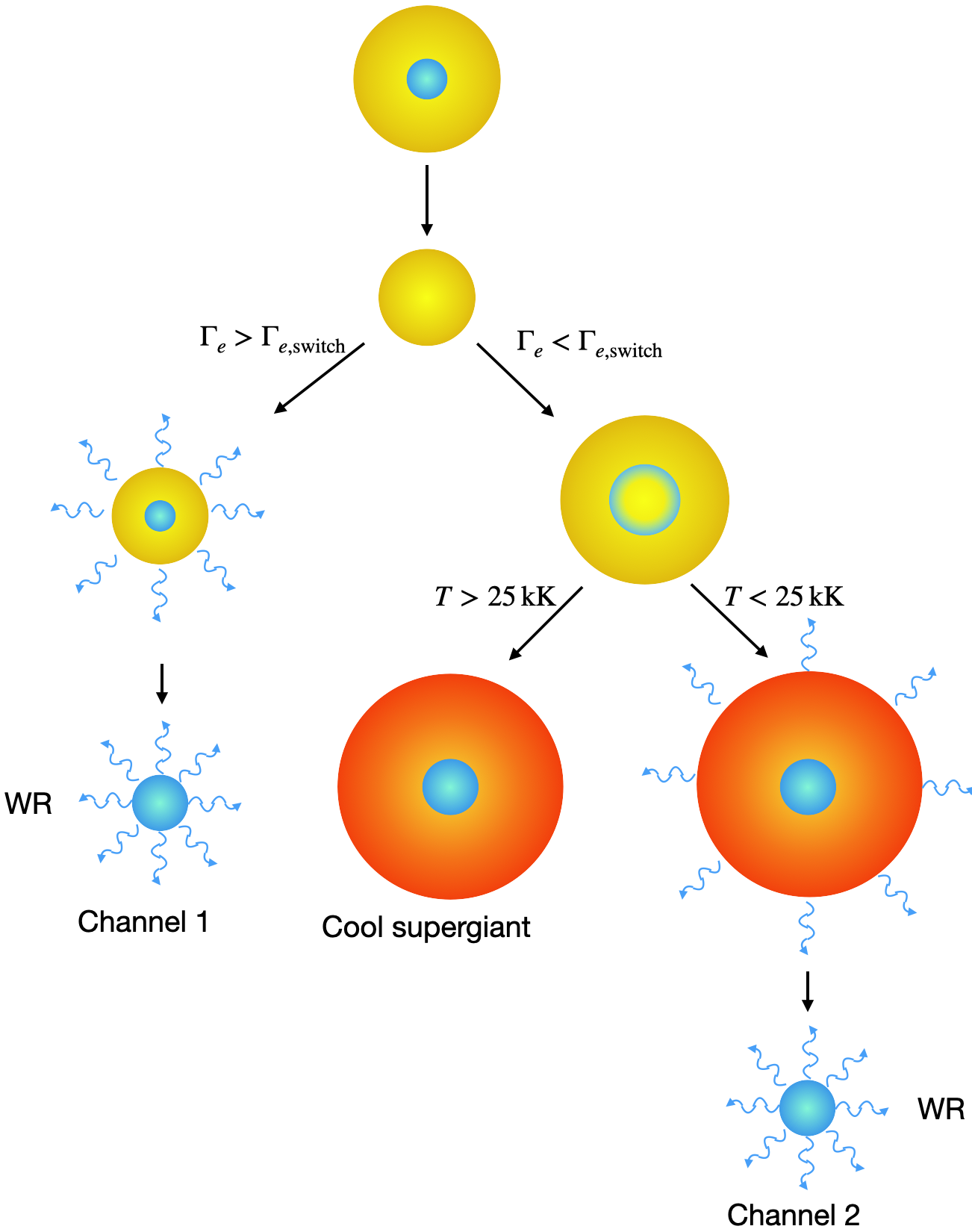}
    \caption{Illustration depicting the possible evolutionary paths of a chemically homogeneous star with initial mass $\gtrsim 20\,\rm M_\odot$ at SMC metallicity. After core H-burning (yellow ball with blue center), the star contracts, $\Gamma_\text{e,switch}$ decreases and $\Gamma_\text{e}$ increases. If $\Gamma_\text{e}>\Gamma_\text{e,switch}$ the star enters the optically thick winds regime (left arrow) and ends up being a WR (channel 1). If $\Gamma_\text{e}<\Gamma_\text{e,switch}$ (right arrow), winds are still optically thin during shell H-burning (yellow ball with blue shell) and the star expands and cools, increasing $\Gamma_\text{e,switch}$. At this stage, there are two possible outcomes: (i) If the star cools enough to cross the bi-stability jump $T_{\rm eff}<25$ kK (right arrow), it may enter the optically thick wind regime and become a WR (channel 2). (ii) If the star starts core He-burning at $T_{\rm eff}>25$ kK (left arrow), $T_{\rm eff}$ rises again, and the star does not cross the bi-stability jump. Optically thick winds are not activated (or are activated too late) and the star ends up being a cool supergiant.} 
    \label{fig:cartoon}
\end{figure}

To qualitatively describe the impact of all these effects, we refer to Figure \ref{fig:cartoon}. While most of the very massive stars simulated in \citetalias{Sabhahit2023} enter the optically thick regime during the main sequence thanks to their high luminosities, lower mass stars (as the ones simulated in this work) can enter the optically thick wind regime only at the end of the main sequence. When core hydrogen burning stops, the star undergoes a rapid phase of contraction, usually referred to as 'hook' in the HR diagram. This rapid contraction phase leads to an increase in the terminal velocity $\varv_\infty$, with a consequent increase in the wind efficiency parameter $\eta$ and a reduction of $\Gamma_\text{e, switch}$. Moreover, if the star is rapidly rotating, this phase is followed by a quick increase in luminosity and $\Gamma_\text{e}$. For these reasons, this is the first good spot for a star with mass $M<100\,\rm M_\odot$ to enter the optically thick regime, provided that its luminosity is high enough and the star is sufficiently compact. We refer to this activation scenario as channel 1. 

If the star does not enter the optically thick regime through channel 1, the contraction phase is followed by the stellar expansion during shell hydrogen burning. During the expansion, both the terminal velocity and the rotation boost factor decrease. Consequently, $\eta$ decreases and $\Gamma_\text{e,switch}$ increases, and the star remains in the optically thin regime. At this point two things may occur: (1) Core He-burning begins when the star is still above the bi-stability jump. (2) The star becomes cold enough ($T_{\rm eff}<25$ kK) to cross the bi-stability jump before the onset of core He-burning. In case (1), core He-burning stops the expansion of the star, thus preventing it from crossing the bi-stability jump. In this case, winds remain optically thin for all the core He-burning phase. Afterward, some stars are able to activate optically thick winds, but that happens too late in their evolution to remove a substantial fraction of the envelope. These stars will end their lives as cool supergiants.

In case (2), instead, the bi-stability jump is crossed before core He-burning. The bi-stability jump reduces the terminal velocity by a factor of $\sim 2$ \citep{Vink2001}. This has a twofold effect. First, since $\eta\propto\dot M_\text{V01}\,\varv_\infty\propto \varv_\infty^{-0.226}$, a reduction in $\varv_\infty$ leads to a slight increase in $\eta$. Second and most important, the ratio $\varv_\text{esc}^2/\varv_\infty^2$ increases by a factor of $\sim 4$, leading to a reduction of $\eta_{\rm switch}$ and, consequently, of $\Gamma_\text{e, switch}$. This drastic reduction of $\Gamma_\text{e, switch}$ can lead to the activation of optically thick winds if the star is luminous enough (see also \citetalias{Sabhahit2023}). This is the second good spot for the star to enter the optically thick wind regime. We call this channel 2. This general behavior is depicted in Figure~\ref{fig:cartoon}. 

\begin{figure}
    \centering
    \includegraphics[width=0.9\linewidth]{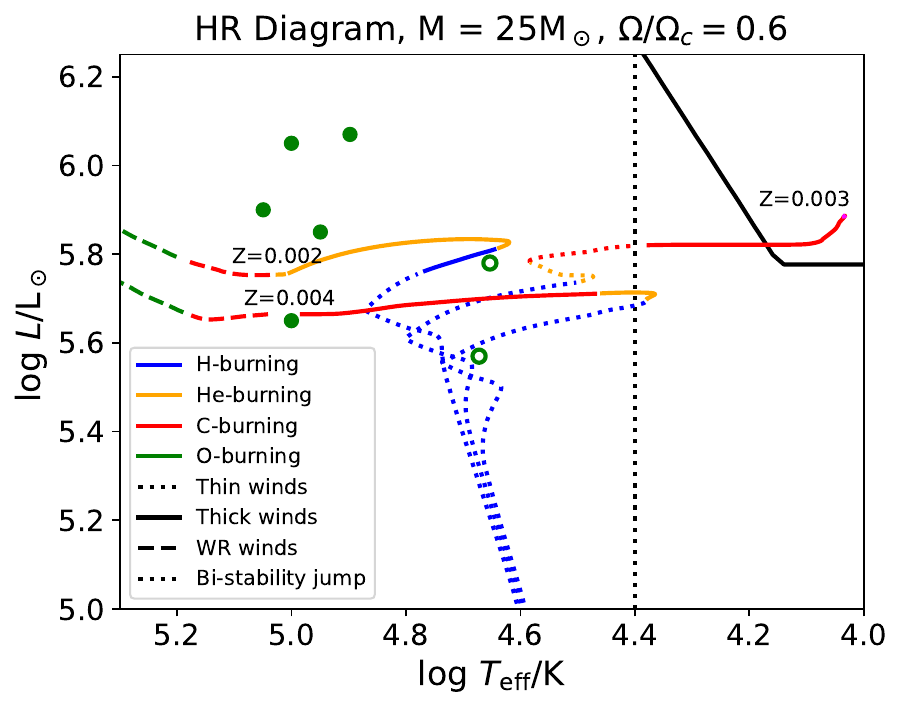}
    \includegraphics[width=0.9\linewidth]{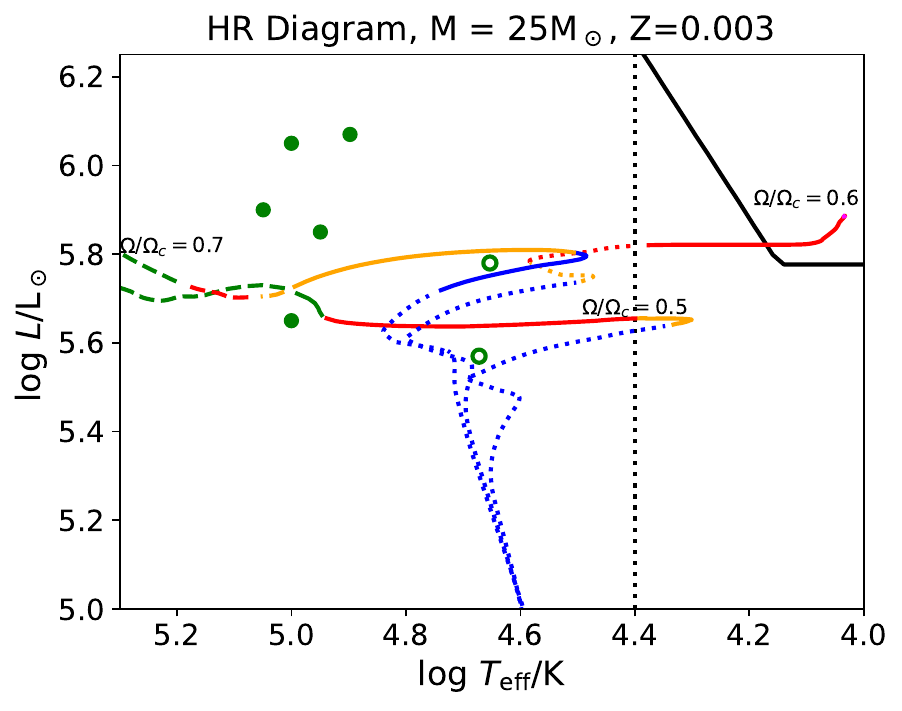}
    \caption{Stellar tracks for a $25$~M$_\odot$ star. The upper panel shows a rotating star with  $\Omega/\Omega_c=0.6$ for three different metallicities $Z=0.002,\,0.003,\,0.004$. The lower panel shows a star with metallicity $Z=0.003$ for three different values of the initial rotation speed $\Omega/\Omega_c=0.5,\,0.6,\,0.7$. The linestyles and color code are the same as in Figure \ref{fig:new_winds}. The vertical dotted black line represents the temperature below which we expect the bi-stability jump. This Figure shows the non-monotonic dependence of mass-loss on metallicity and rotation. Low metallicity ($Z=0.002$) favors the activation of optically thick winds through channel 1, because the star is more luminous and compact. Higher metallicity ($Z=0.004$) can also  activate optically thick winds through channel 2. Stars in the intermediate metallicity case ($Z=0.003$), instead, fail to activate optically thick winds soon enough and end up as cool supergiant stars. The same trend happens for rotation.}
    \label{fig:metallicity_dependence}
\end{figure}

To see these effects in more detail, the upper panel of Figure \ref{fig:metallicity_dependence} shows three different tracks for a star with initial mass $M=25\,\rm M_\odot$ and initial rotation $\Omega/\Omega_c=0.6$ at three different metallicities: $Z=0.002$, $0.003$, and $0.004$. As said above, lower metallicity tracks have higher luminosities and temperatures. The $Z=0.002$ track is luminous enough to activate optically thick winds right after the contraction phase at the end of the main sequence (channel 1). 

The $Z=0.003$ track, instead, is not luminous enough to activate optically thick winds at the end of core H-burning and evolves toward cooler temperatures. At $\log (T_{\rm eff}/{\rm K})\simeq 4.5$, core He-burning starts and the star contracts again. It is only later, during carbon burning, that the star crosses the bi-stability jump and activates optically thick winds. This is too late to peel off the hydrogen-rich envelope, and the star ends up as a cool supergiant, eventually crossing the HD limit. 

The $Z=0.004$ track is even cooler and not luminous enough to activate optically thick winds through channel 1. It evolves toward cooler temperatures and crosses the bi-stability jump before core He-burning. In our model, this leads to the activation of optically thick winds through channel 2, and the star moves to the hot region of the HR diagram. 

Therefore, the metallicity dependence is not monotonic. At lower metallicities, optically thick winds activate because of the high luminosity and hot temperatures (channel 1). At higher  metallicities, optically thick winds activate because cooler stars can reach the bi-stability jump before core He-burning (channel 2). At intermediate metallicities, instead, the stars struggle to enter the optically thick wind regime, as they are not luminous enough to enter the optically thick wind regime at the end of the main sequence, but they are too hot to cross the bi-stability jump before the onset of core He-burning.

A similar behavior applies to rotation, with higher rotations corresponding to more luminous and hotter stars and lower rotations to less luminous and cooler ones. This is shown in Figure \ref{fig:metallicity_dependence} (lower panel) where we display three different tracks for a star with initial mass $M=25\,\rm M_\odot$ at metallicity $Z=0.003$, with initial rotation $\Omega/\Omega_\text{c}=0.5,\, 0.6,$ and 0.7. The fastest rotating star in our example ($\Omega=0.7\,{}\Omega_\text{c}$) activates optically thick winds through channel 1, while the one rotating more slowly ($\Omega=0.5\,{}\Omega_\text{c}$) becomes a WR through channel 2. The star with intermediate rotation velocity ($\Omega=0.6\,{}\Omega_\text{c}$), instead, is not able to activate any of the two channels. However, we note that an even slower rotation ($\Omega<0.5\,{}\Omega_\text{c}$ in our example) will not activate optically thick winds even if the star crosses the bi-stability jump before core He-burning. This happens because the star needs enough rotation to enter the chemically homogeneous evolution regime; otherwise its luminosity would be too low to activate optically thick winds.

A similar dependence is found for the overshooting parameter, which has a very similar effect to rotation (see Section \ref{sec:overshooting}). A more detailed analysis of the behavior of $\Gamma_\text{e}$, $\Gamma_\text{e, \rm switch}$, $\eta$ and $\eta_{\rm switch}$ can be found in Appendix A. The outcome of all the simulations with different masses, metallicities, and initial rotations can be found in Table \ref{fig:tables} in Appendix B.  

As a final remark, we note that while channel 1 implies that the star enters the optically thick regime due to its compactness and luminosity, channel 2 heavily relies on the bi-stability jump and thus has to be considered more uncertain. The existence and strength of the bi-stability jump have intensely been debated over the last few years. We refer to Appendix~\ref{app:jump} for additional discussion about the bi-stability jump.

\subsection{WR evolution and their properties}\label{sec:properties}
We have shown that even stars with mass down to $M\sim 20$~M$_\odot$ can develop optically thick winds if their initial rotation is high enough. Winds peel off the stellar envelope and make the star move to the hot part of the HR diagram. We now describe additional properties of these stars and assess whether they are compatible with single WRs in the SMC. We consider different properties of our simulated stars: the time they spend in a given region of the HR diagram, their surface abundances, their average rotation velocity, and the transformed mass-loss rate $\dot M_\text{t}$. Given the uncertainties on the bi-stability jump mentioned above, we take as reference the case at metallicity $Z=0.002$ and $\Omega/\Omega_c=0.6$, where all stars become WRs through channel 1.

\subsubsection{Time evolution on the HR diagram}
To calculate the time spent by our simulated stars in each region of the HR diagram, we first defined the quantity
\begin{equation}
\varv_{\rm HRD}=\sqrt{\left|\frac{d\log L}{dt}\right|^2+\left|\frac{d\log T}{dt}\right|^2},
\end{equation}
which represents the ``transition'' velocity of the star on the HR diagram and has units of dex yr$^{-1}$. The inverse of $\varv_{\rm HRD}$, $\tau_{\rm HRD}\equiv 1/\varv_{\rm HRD}$, with units of yr dex$^{-1}$, measures the time spent by the star to move by 1 dex in the HR diagram. 

\begin{figure}
    \centering
    \includegraphics[width=0.9\linewidth]{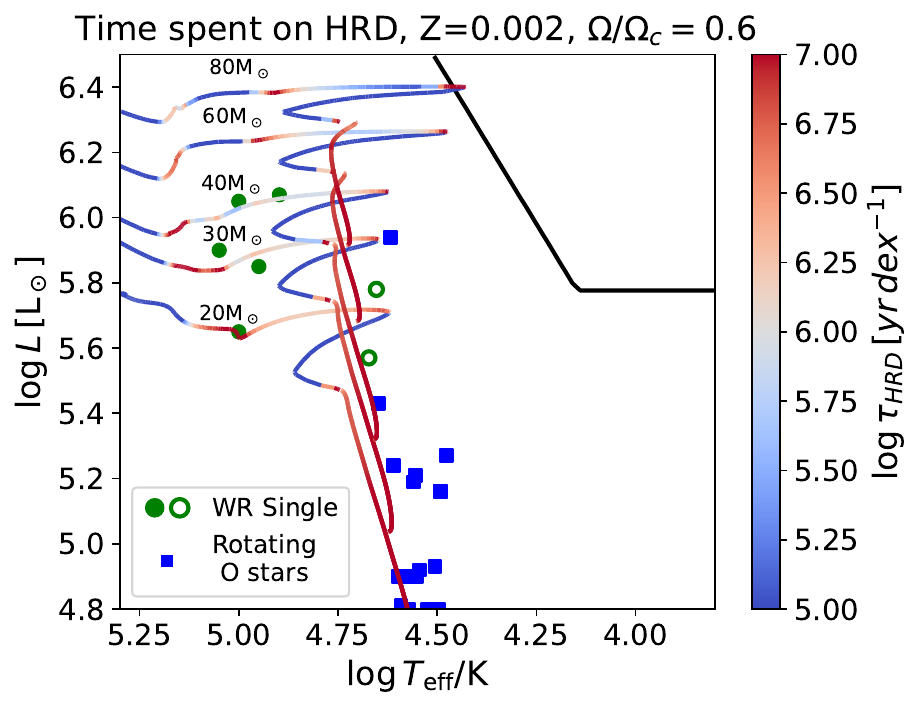}
    \includegraphics[width=0.9\linewidth]{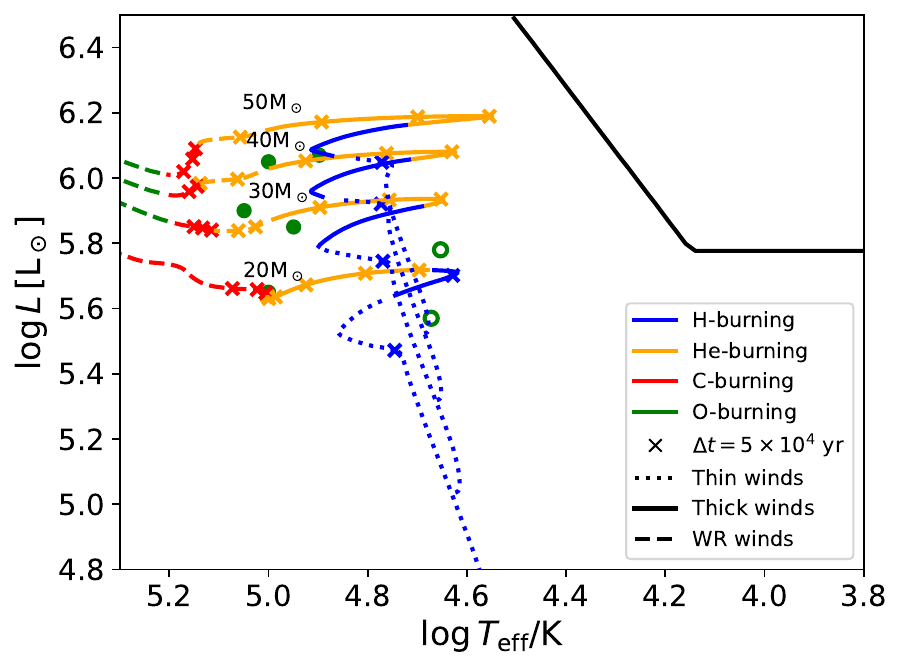}
    \caption{Time spent on different regions of the HR diagram for stars with mass $M=20,\,30,\,40,\,60$,\,and $80$~M$_\odot$ at $Z=0.002$ and $\Omega/\Omega_c=0.6$. In the upper panel, the color code represents the quantity $\tau_\text{HRD}$, in yr $\rm dex^{-1}$, which is the time spent to move by $1\,\rm dex$ in the HR diagram. Green circles are WRs in the SMC ,and blue squares are O-type stars in the SMC with $\varv\,\sin i>200$ km/s from \citep{Mokiem2006, Bouret2013, Bouret2021, Ramachandran2019, Dufton2019, Rickard2024, Backs2024}. In the lower panel, the color code represents different burning stages: core H-burning (blue), He-burning (orange), carbon burning (red), and oxygen burning (green), while the line styles represent different wind regimes: dotted for optically thin winds, solid for optically thick winds, dashed for WR-type winds. Cross markers are spaced by $\sim 5\times 10^4\,\rm yr$ each. After the main sequence, stars spend most of their time at the turn-off point, where they start to peel off their H-rich envelope and move blueward, and in the hot WR phase, at $4.9<\log (T_{\rm eff}/{\rm K})<5.2$.}
    \label{fig:time}
\end{figure}

The upper panel of Figure \ref{fig:time} shows stellar tracks on the HR diagram for $Z=0.002$ and $\Omega/\Omega_c=0.6$ colored by $\tau_{\rm HRD}$. The lower panel displays the same tracks with a cross marker every $\sim 5\times 10^4\,\rm yr$, colored by burning stage. It is clear that all stars spend most of their post main sequence lifetime in two regions: (1) near the turn-off point, when they start losing their envelope and moving blueward, and (2) during the hot WR stage. The exact effective temperatures of these stages vary with stellar mass.

For lower mass tracks ($20-40\,\rm M_\odot$), region (1), corresponding to early core He burning, is at $\log (T_{\rm eff}/\rm K)\gtrsim 4.6$, while region (2) is at $\log (T_{\rm eff}/\rm K)\sim 5-5.1$ during mid/late core He-burning. We accurately reproduce the position of the lowest luminosity observed WR at $\log (T_{\rm eff}/\rm K)\sim 5$, where the $M=20$~M$_\odot$ track spends most of its post main sequence time. Tacks with $30-40\,\rm M_\odot$, instead, tend to have slightly higher effective temperatures than observed stars, by $\sim 0.1\,\rm dex$.

The $20-40,\rm M_\odot$ tracks intersect the HR positions of the five observed hot WRs during mid-to-late core He burning. Although the ``WNh'' label is often associated with core hydrogen-burning stars having a WN-type appearance, we can exclude this possibility here. Extending the main sequence to effective temperatures as high as $\log (T_{\rm eff}/\rm K)\sim 5$ would require fully chemically homogeneous evolution, which in turn would imply very low hydrogen abundances throughout the star, including the surface. This is at odds with their observed WNh classification. Alternatively, assuming high hydrogen surface abundances would reflect a large, hydrogen-rich shell, and would imply a stellar mass that is so high that the resulting $L/M$-ratio would not yield a WR-type spectrum \citep[cf.][]{Shenar2020,Sander2024a,Sander2024b}.

Overall, we interpret the five hot WNh in the SMC as being in a similar evolutionary stage as WN-early stars observed in the Milky Way or the Large Magellanic Cloud except that they retain some surface hydrogen, likely due to weaker winds, which preserves their WNh spectral features. Instead, the two cooler WNh stars can be interpreted as late main sequence or early core He burning objects. From Figure \ref{fig:time}, the 2:5 ratio of cold to hot WNh roughly corresponds to the relative time stars spend in these two HR regions. However, we warn that this ratio may substantially depend on stochastic fluctuations and on the details of the star formation history of the SMC, which we do not consider in this work. Thus, we refrain from drawing further conclusions based on such number counts.

For higher mass tracks ($60-80\,\rm M_\odot$), region (1) is at $\log (T_{\rm eff}/\rm K)\gtrsim 4.4$ and region (2) at $\log (T_{\rm eff}/\rm K)\sim 4.9$. These high-mass stars also spend some time at higher temperatures, $\log (T_{\rm eff}/\rm K)\sim 5.2$, corresponding to carbon burning stage. The absence of observed WRs in the SMC in this mass regime is not surprising given the general deficiency of very luminous massive stars in the SMC and it is likely the result of its star formation history \citep{Ramachandran2019,Schootemeijer2021}.

In the upper panel of Figure~\ref{fig:time}, we also report data for O-type stars in the SMC with high projected rotational velocity ($\varv\,\sin i>200$ km/s). Most of these stars cluster around the main-sequence evolutionary tracks for $20, 30, 40\, M_\odot$ stars, where they spend most of their lifetime. However, one rapidly rotating O-type star lies right after the hook, at $\log (L/L_\odot)\lesssim 6$, where the star temporarily stalls before evolving toward the hotter region of the HR diagram.

The results shown here are for $Z=0.002$, where we find WRs to be mainly produced by channel 1. In Appendix \ref{app:ch2}, we present the same analysis for $Z=0.004$, where channel 2 is the primary WR formation channel. From Figure \ref{fig:time_004}, it is evident that hot WRs formed through channel 2 are more evolved than those formed through channel 1. Specifically, the channel 2 tracks are already in the core C burning phase when they reach the observed HRD location.

\subsubsection{Surface abundances}
\begin{figure}
    \centering
    \includegraphics[width=0.9\linewidth]{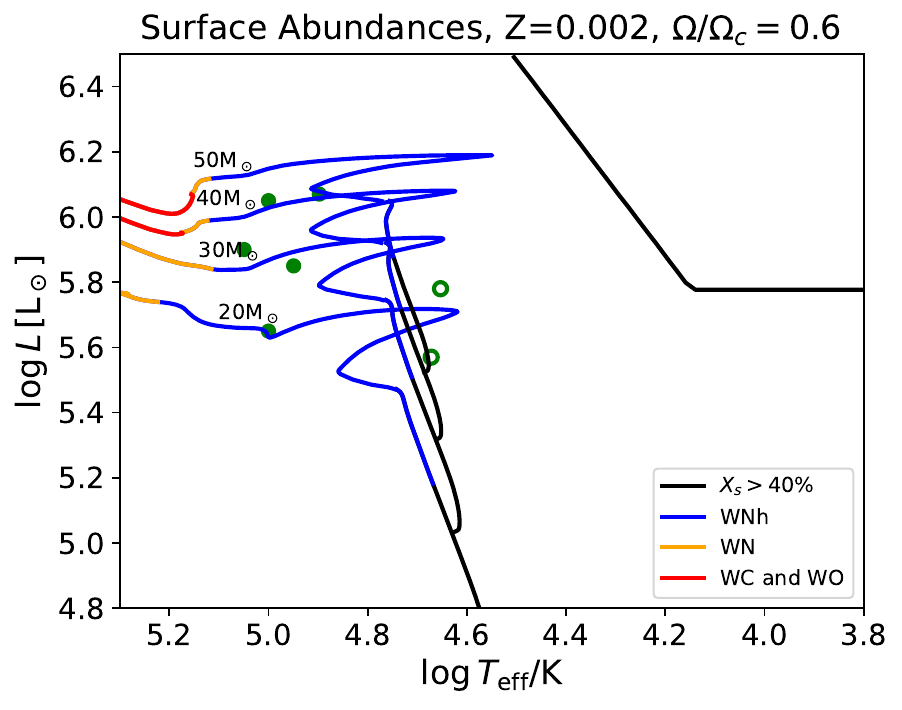}
    \includegraphics[width=0.9\linewidth]{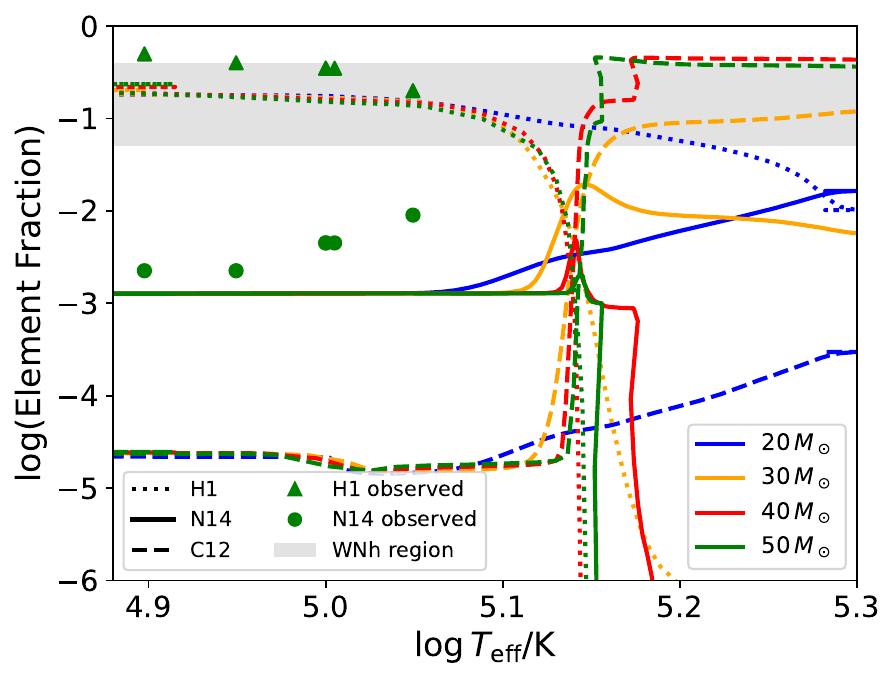}
    \caption{Upper panel: Classification of the star along the HR diagram: O-type (black), WNh (blue), WN (orange), and WC and WO (red). Lower panel: Surface abundances as a function of the effective temperature, $T_{\rm eff}$. Line styles are for different elements: hydrogen (dotted line), $^{14}$N (solid line), $^{12}$C (dashed line). Different colors are for different initial masses: 20 (blue), 30 (orange), 40 (red), and 50~M$_\odot$ (green). Green triangles (circles) represent hydrogen (nitrogen) abundances for the five hottest WRs observed \citep{Hainich2015}. The gray horizontal band covers the $\sim 5-40\%$ range; if the hydrogen abundance is inside this region, the star is considered a WNh.}
    \label{fig:abundances}
\end{figure}

Figure \ref{fig:abundances} shows the surface abundances at $Z=0.002$ and rotation $\Omega/\Omega_c=0.6$. The upper panel shows the evolution on the HR diagram and the color represents the WR type. We define as WNh (or slash) all stars with surface H abundance $X_s$ in the range $5-40\%$ and in the optically thick winds regime. The star is a WN if the surface H abundance drops below $5\%$ and the $^{12}$C or $^{16}$O surface abundances are $<20$\%. If the $^{12}$C ($^{16}$O) surface abundance is  $>20\%$, we label the star as WC (WO). WN, WC, and WO stars are also called classic WRs.

The WNh phase begins at the end of the main sequence, when thick winds are activated, and lasts until $\log (T_{\rm eff}/{\rm K})\gtrsim 5.1$. Afterward, surface hydrogen is completely removed by the wind, and the star has a WN phase before becoming carbon or oxygen dominated. The classic WR phase occurs at very high temperatures. Lower mass stars $M\leq 30\,\rm M_\odot$, do not reach the carbon or oxygen dominated phase by the end of our simulation. 

The bottom panel shows the evolution of surface abundances of hydrogen, $^{14}$N and $^{12}$C for different stellar masses, as a function of the effective temperature. Since the evolution of $T_{\rm eff}$ is not monotonic, we plot only the $\log (T_{\rm eff}/\textrm{K})\gtrsim 4.9$ regime, for the sake of clarity. In this regime, $T_{\rm eff}$ always increases with time. All stars feature $\sim 20\%$ surface hydrogen for a huge part of their evolution, until $\log (T_{\rm eff}/\textrm{K})\sim 5.1$, corresponding to $\sim 97\%$ of the star lifetime. During this phase, they are classified as WNh, in agreement with observational data, although the observed WNh in the SMC show a wider range of surface hydrogen abundances, $\sim 20-50\%$. 

This discrepancy was already discussed by \citet{Schootemeijer2018}, who showed that reproducing the effective temperatures of the hot SMC WRs through chemical homogeneous evolution leads to underestimation of their surface hydrogen content. In our scenario, chemical homogeneous evolution is instrumental for the activation of thick winds, but it is not the main driver of the transition to the hot side of the HR diagram. Because lower initial rotations are sufficient to trigger optically thick winds and envelope removal, our models retain somewhat higher surface hydrogen fractions than in a purely homogeneous scenario. While they still fall short of reproducing surface H abundances as high as $\sim 50\%$, this is an improvement compared to a purely chemically homogeneous evolution scenario. Moreover, Appendix~\ref{app:ch2} shows that models at $Z=0.004$, which evolve through channel 2, are in better agreement with the observed  surface hydrogen abundance. At $\log (T_{\rm eff}/\textrm{K})\gtrsim 5.1$, hydrogen is rapidly eroded by winds, which progressively expose the internal structure and, when surface hydrogen drops below $0.05\%$, we classify the star as a WN. 

The nitrogen abundance increases quite rapidly during the main sequence as a result of the CNO cycle, rising from an initial value of $\sim 10^{-4}$ to an equilibrium value of $\sim 10^{-3}$ (see Appendix \ref{app:profile}). This enhancement is distributed nearly homogeneously throughout the star owing to rotational mixing. Once core He-burning begins, nitrogen in the core is quickly depleted, while the surface abundance remains unchanged. At $\log (T_{\rm eff}/\textrm{K})\gtrsim 5.1$, there is a further enhancement of surface nitrogen. This is due to the combined effect of envelope hydrogen burning and winds exposing the interior of the star. Afterward, the nitrogen surface abundance declines as it is eroded by winds, which expose the nitrogen depleted core of the star. The rise and fall of surface nitrogen occur more rapidly in more massive stars, due to their stronger winds. The predictions of our models are in agreement with the two data points at $\log (T_{\rm eff}/\textrm{K})<5$, featuring a surface $^{14}$N $\sim 2\times 10^{-3}$.

We note that here we show only the lowest metallicity case $Z=0.002$. For an initial metallicity of $Z=0.004$, the models predict an equilibrium nitrogen abundance of $\sim 2\times 10^{-3}$, in good agreement with the data (see Appendix \ref{app:ch2}). Observed WRs in the SMC, however, display surface nitrogen enrichment already at $\log (T_{\rm eff}/\textrm{K})\gtrsim 5$, whereas in our models the enrichment becomes visible only beyond $\log(T_{\rm eff}/\mathrm{K}) \gtrsim 5.1$. Nevertheless, our models do produce a nitrogen “bump” around $\log(T_{\rm eff}/\mathrm{K}) \sim 5$, but this enhancement is confined to layers just beneath the surface and only becomes observable once stellar winds peel off the outermost material (Appendix~\ref{app:profile}).

Overall, we attribute this discrepancy in the timing of surface nitrogen enrichment to the poor understanding of the rotational mixing processes. As discussed in Section~\ref{sec:methods}, we adopt the \citet{Heger2000} calibration of the mixing efficiency parameters $f_c$ and $f_\mu$, based on nitrogen abundances in solar metallicity terminal-age main-sequence stars. Alternative calibrations \citep[e.g.,][]{Brott2011, Costa2019a} may be more appropriate for SMC metallicities. However, a detailed tuning of these parameters is beyond the scope of the present work.

Another possible explanation for the mismatch in the effective temperature of the nitrogen enrichment could be traced back to the radius problem. Since simulated WRs tend to be at a higher $T_{\rm eff}$ than the observed ones by $\sim 0.1\rm dex$ (see also Figure \ref{fig:time}), a shift of the data to higher temperatures by this amount would allow the effective temperature of the nitrogen enrichment to be matched.

Carbon surface abundance follows a trend similar to that of nitrogen. At $\log(T_{\rm eff}/\mathrm{K}) < 5.1$, it remains nearly constant at $\sim2\times10^{-5}$, independent of stellar mass. Beyond this point, the surface carbon abundance increases as stellar winds expose deeper layers. Oxygen features a very similar trend as carbon, but it is not shown in the figure for clarity. By the end of the evolution, more massive stars ($40-50\,\rm M_\odot$) reach $\gtrsim 80\%$ in $^{12}$C + $^{16}$O abundance, being classified as WC or WO. In contrast, the $20$ and $30~\rm M_\odot$ tracks have not reached the WC or WO phase even at temperatures as high as $\log (T_{\rm eff}/{\rm K})\sim 5.3$, with their surface carbon abundance still increasing with temperature.

\subsubsection{Rotation}
\begin{figure}
    \centering
    \includegraphics[width=0.9\linewidth]{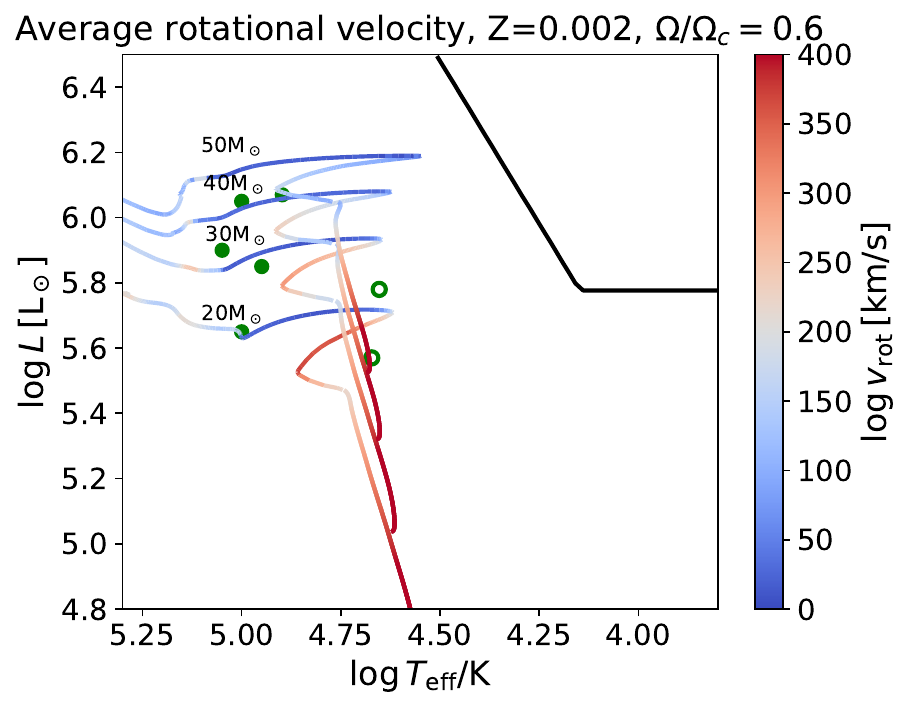}
    \includegraphics[width=0.9\linewidth]{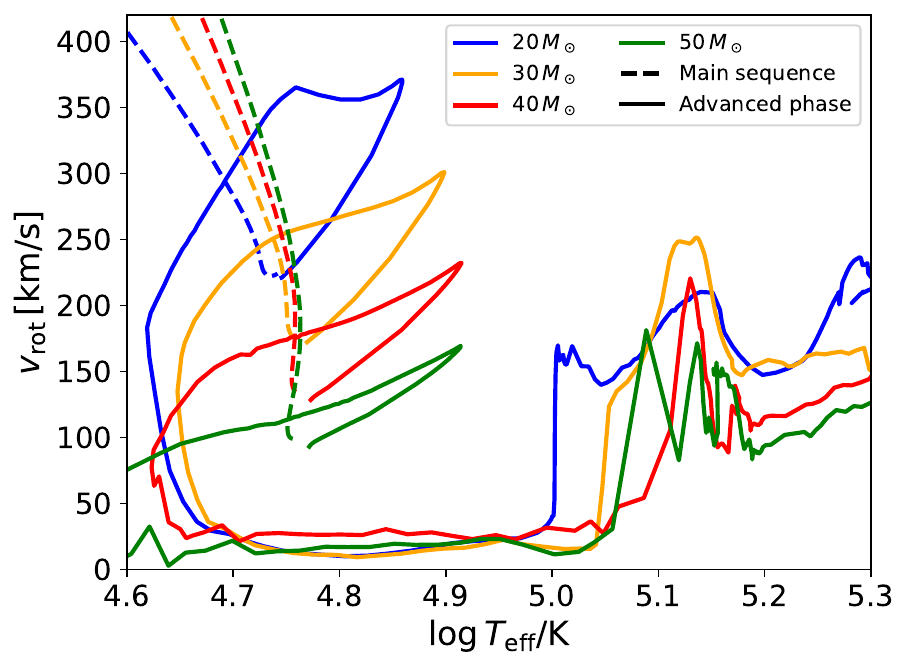}
    \caption{Upper panel: Average rotational velocity along the HR diagram (color code). Lower panel: Evolution of the average rotational velocity as a function of the $T_{\rm eff}$. The color code is the same as in Figure \ref{fig:abundances}. Dashed lines represent the main sequence evolution, solid lines the advanced phases. The velocity drops during the main-sequence evolution. This is followed by the hook at the end of the main sequence, when the velocity quickly rises due to post main sequence contraction, and by a subsequent drop. In the late phases of the evolution $\log (T_{\rm eff}/{\rm K})\gtrsim 5.05$, the  velocity rises again due to winds peeling-off the outer layers of the star and exposing the rotating core.}
    \label{fig:vsurf}
\end{figure}

Figure \ref{fig:vsurf} shows the evolution of the rotation velocity of the star in the HR diagram (top) and as a function of $T_{\rm eff}$ (bottom). The star needs a high initial rotation to develop chemically homogeneous evolution \citep{Brott2011}. However, after approximately half of the stellar lifetime, the velocity drops below $400$ km/s, reaching values more similar to those observed for the bulk of the O-type star population. The drop is faster for more massive stars, with the 50 M$_\odot$ star that can reach a rotation speed as low as $\sim 100$ km/s during the main sequence. After the end of core H-burning the star contracts and the rotational velocity increases, before dropping to almost zero due to the activation of optically thick winds which remove angular momentum. Finally, toward the end of the evolution, when $\log (T_{\rm eff}/{\rm K})\gtrsim 5.05$ ($\gtrsim 97\%$ of the star lifetime), the rotational velocity increases again to values $\sim 100-250$ km/s due to winds exposing the rotating core of the star. 

While we do not have measures of the rotational velocity of WRs, there are some broad upper limits on $\varv\sin i$ reported in the literature. From the shape of narrow absorption/emission lines \citet{Martins2009} find $\varv\sin i<50\,{\rm km/s}$ for WRs in the SMC. \citet{Hainich2015} revised this estimate by considering that spectral lines of WRs are formed in the stellar wind, obtaining less stringent velocity constraints of $\varv\sin i<200\,{\rm km/s}$. Using spectropolarimetry, \citet{Vink2017b} found a quite low $\sim 10\%$ incidence of line effects on WRs in the LMC and detected only one line effect in the SMC WRs. This implies that observed WRs should feature moderate rotational velocities. Our tracks are compatible with these upper limits. They start to rise to $\varv_{\rm rot}>100\,{\rm km/s}$ when the temperature $\log (T_{\rm eff}/{\rm K})\gtrsim 5.05$ due to the exposition of the core. Since all observed WRs in the SMC have $\log (T_{\rm eff}/{\rm K})< 5.05$, they are still in the phase with $\varv_{\rm rot}<50\, {\rm km/s}$, consistent even with the most stringent upper limits.

\subsubsection{mass-loss}
\begin{figure}
    \centering
    \includegraphics[width=1\linewidth]{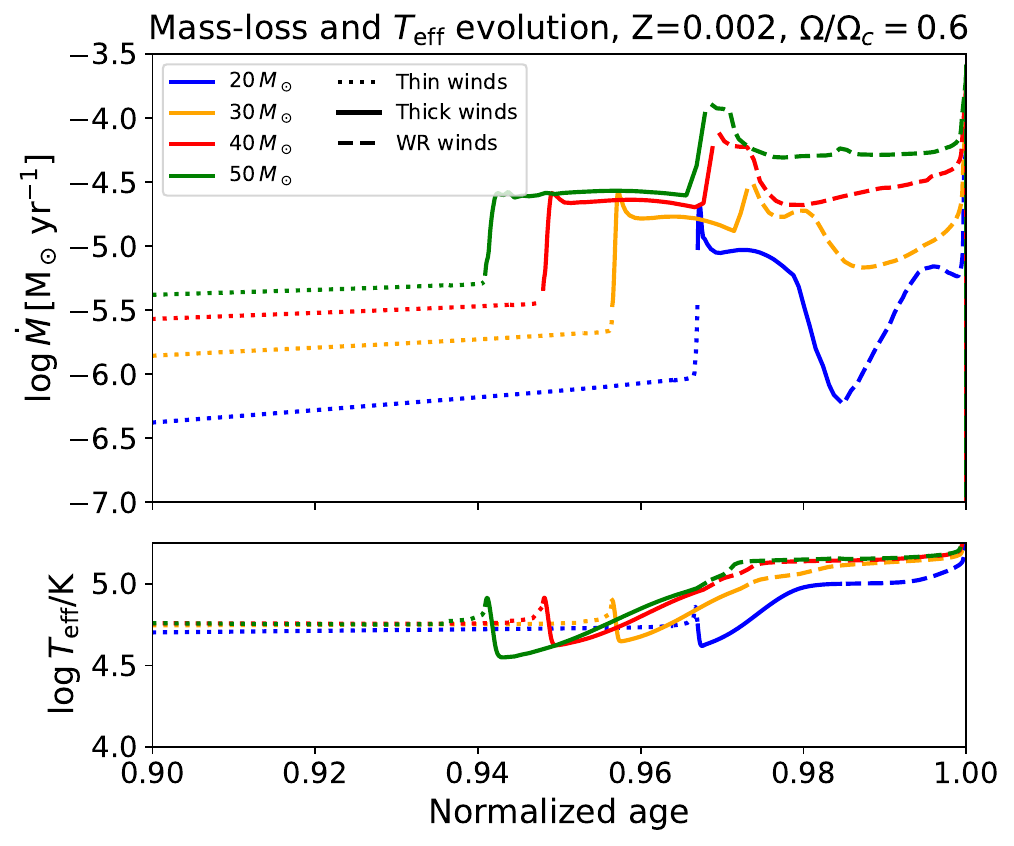}
    \caption{mass-loss, $\dot M$, (upper panel) and $\log (T_{\rm eff}/{\rm K})$ (lower panel) as a function of normalized stellar age. The color code is the same as in Figure \ref{fig:abundances}. Line styles represent different wind regimes: dotted for optically thin winds, solid for optically thick winds, dashed for WR-type winds.}
    \label{fig:mass-loss}
\end{figure}

Figure \ref{fig:mass-loss} shows the evolution of mass-loss and effective temperature as a function of the normalized stellar age, defined as the fraction of the total stellar lifetime. During the main sequence, mass-loss is driven by optically thin winds, whose strength gradually increases as luminosity rises. At the end of hydrogen burning, the star contracts, both effective temperature  and luminosity increase, and optically thick winds are activated through channel 1. This correspond to an enhancement of mass-loss of about one order of magnitude. In the optically  thick-wind regime, mass-loss remains approximately constant or shows a slight decline due to the modest luminosity decrease. At $\log (T_{\rm eff}/{\rm K})>5$ WR-type winds are activated, and the mass-loss rate can exhibit sharp variations owing to its steep dependence on $\Gamma_\text{e}$.

We computed the ``transformed mass-loss rate'' \citep{Graefener2013}, which is defined as
\begin{equation}
\dot M_\text{t}=\dot M\,\sqrt{D}\,\left(\frac{1000\, {\rm km}\,{}{\rm s}^{-1}}{\varv_\infty}\right)\,\left(\frac{10^6\,{\rm L}_\odot}{L}\right)^{3/4},
\label{eq:dotMt}
\end{equation}
where $D$ is the clumping factor assumed to be $D\simeq 10$ when calculating it from the evolution models. The quantity  $\dot{M}_\text{t}$ provides the (unclumped) mass-loss rate a star would have if it had a luminosity of $10^6\,L_\odot$ and a wind terminal velocity of $1000\,\mathrm{km}\,\mathrm{s}^{-1}$. Stars with the same temperature and $\dot{M}_\text{t}$ have approximately the same normalized line spectrum \citep[see also][]{Schmutz1989}. As the ``raw'' mass-loss rate increases considerably with the luminosity, $\dot{M}_\text{t}$ is a better quantity to compare the wind strength of stars with different luminosities and to judge whether our tracks not only lie in the same HR diagram position as the single WRs of the SMC, but also show a mass-loss rate approximately compatible with the observed strength.

\begin{figure}
    \centering
    \includegraphics[width=0.9\linewidth]{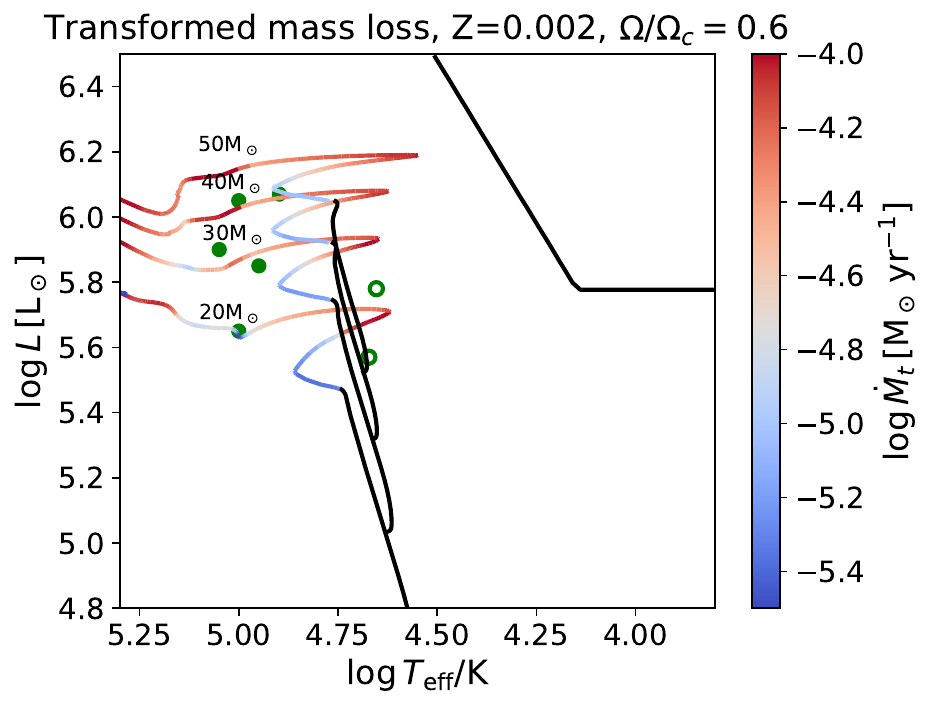}
    \includegraphics[width=0.9\linewidth]{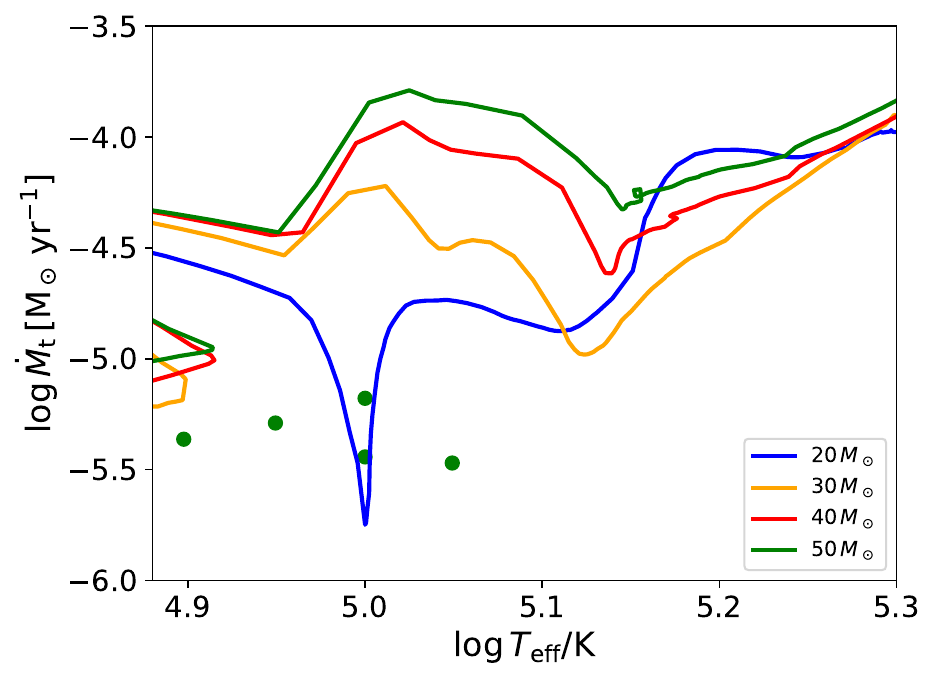}
    \caption{Upper panel: Transformed mass-loss, $\dot M_t$, along the HR diagram track (color code). Lower panel: Evolution of $\dot M_t$ as a function of $T_{\rm eff}$. The color code is the same as in Figure \ref{fig:abundances}. The green points are computed from the observed $\dot M$, $L$, and $v_\infty$ from \citet{Hainich2015}.}
    \label{fig:transformed_massloss}
\end{figure}

Figure \ref{fig:transformed_massloss} shows the evolution of $\dot{M}_\text{t}$ in the HR diagram (top) and as a function of the effective temperature (bottom). With the activation of the optically thick winds, $\dot M_\text{t}$ rises to values $\sim 10^{-4.5}$~M$_\odot/$yr. This is followed by a decrease, more evident for less massive stars, due to a drop in the applied mass-loss recipe. Afterward, when $T_{\rm eff}>10^5$~K, $\dot M_t$ rises again due to the activation of WR-type winds \citep{Vink2017,Sander2020}. This is then followed by a shallow decrease, due to a decline in $\Gamma_\text{e}$, and by another increase at $\log (T_{\rm eff}/{\rm K})\gtrsim 5.1$ due to the activation of core C burning, which increases the luminosity and thus $\Gamma_\text{e}$. The decline and rise at $\log (T_{\rm eff}/{\rm K})=5$ is very strong for the $20\, \rm M_\odot$ star because of two reasons: (i) the WR-type winds are extremely sensitive to $\Gamma_\text{e}$ \citep[see][]{Sander2020a}, and thus a slightly lower $\Gamma_\text{e}$ implies much weaker winds. (ii) The activation of core C burning for the $20\,\rm M_\odot$ star occurs exactly at $\log( T_{\rm eff}/{\rm K})=5$, causing the steep rise. We note that we do not include the temperature dependence on $\dot{M}$ from \citet{Sander2023}, which could soften this mass-loss increase as $\dot{M}$ is expected to decline for more compact stars.

The observed $\dot M_\text{t}$-values are computed from Eq.\,\eqref{eq:dotMt} using the values of $\dot M$, $L$, and $\varv_\infty$ reported by \citet{Hainich2015}. Our tracks, especially the ones for $M\geq 30\,\rm M_\odot$, overestimate the value of $\dot M_\text{t}$ by a factor $\sim 10$, predicting a mass-loss that is too high, at least during the WR phase. However, as already mentioned above, WR-type winds feature a steep dependence on $\Gamma_\text{e}$. At the observed $T_{\rm eff}$, our tracks have $\Gamma_\text{e}\sim 0.6$, while a value of $\Gamma_\text{e}\sim 0.5$ would be more than sufficient to be in agreement with observations. Still, such an overprediction of the mass-loss should be carefully regarded for future refinements of our model, and points toward the direction of having less aggressive winds at least during the WR phase. This would also help in reproducing the higher surface hydrogen abundances of the WNh stars in the SMC (see Figure \ref{fig:abundances}).

\section{Discussion}\label{sec:discussion}

\subsection{Overshooting}\label{sec:overshooting}
\begin{figure*}
    \centering
    \includegraphics[width=1.\linewidth]{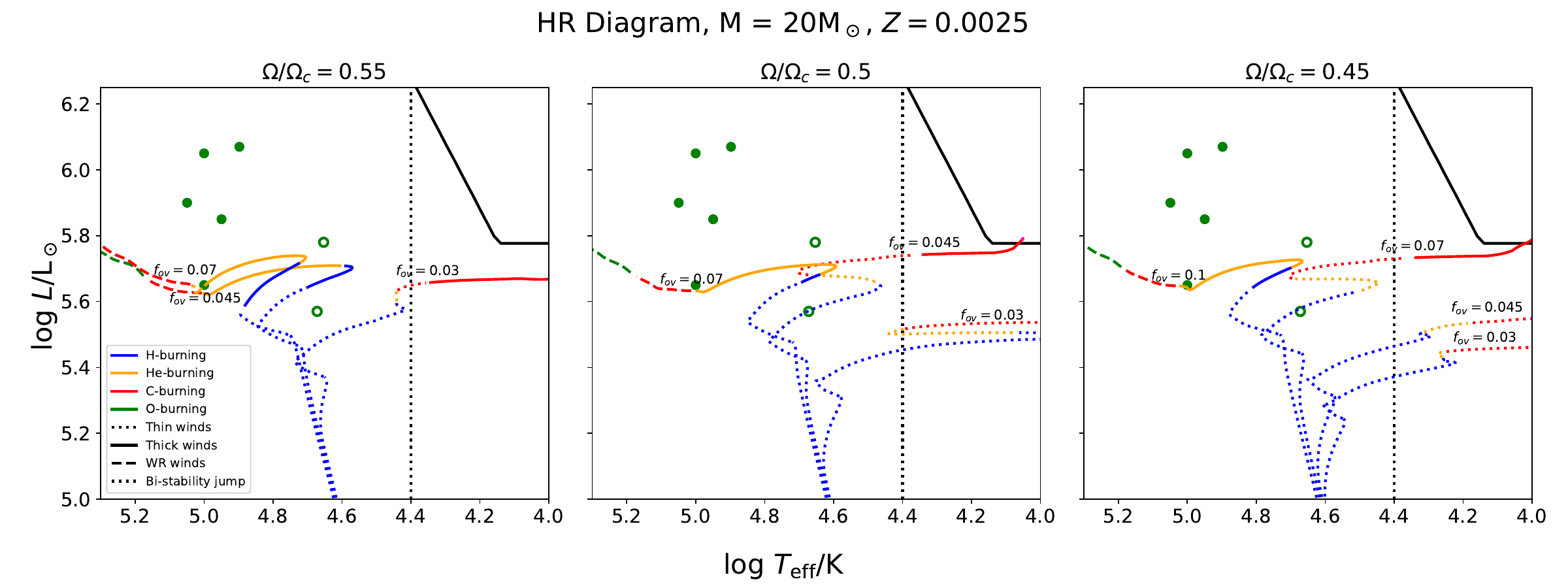}
    \caption{Stellar tracks for a star of initial mass $M=20$ M$_\odot$ at $Z=0.0025$ for initial rotation speed $\Omega/\Omega_c=0.55$ (left), $0.5$ (middle), and $0.45$ (right) and overshooting parameter $f_{\rm ov}=0.03,$ 0.045, 0.07, and 0.1 (right panel) during the main sequence. The effect of overshooting is almost degenerate with that of rotation. A slight increase of $f_{\rm ov}$ corresponds to a reduction of the rotation threshold needed for self-stripping.}
    \label{fig:overshooting}
\end{figure*}

The results shown in Sections \ref{sec:first_results} and \ref{sec:properties}, are obtained with an overshooting parameter  $f_\text{ov}=0.03$ during the first step of the simulation, corresponding to the main sequence, and $f_\text{ov}=0.01$, during the advanced phase. We choose these values according to \citetalias{Sabhahit2023}. However, while these values are expected for stars with mass $\lesssim 5\,\rm M_\odot$ \citep{Herwig2000}, there are some hints that the overshooting parameter might be higher $\gtrsim 0.05$ for stars with larger masses \citep{Vink2010, Higgins2019}. Since overshooting enhances mixing, it has an effect similar to rotation, making the star more luminous and compact, and facilitating the activation of optically thick winds. In particular, for higher $f_\text{ov}$ values a lower initial rotation is required to enter chemically homogeneous evolution and to activate thick winds. This is shown in Figure \ref{fig:overshooting}. The three panels represent the stellar tracks of a $20$ M$_\odot$ star at $Z=0.0025$, with three initial rotation values. The different tracks are for $f_\text{ov}=0.03,\,0.045,\,0.07$ during the main sequence phase, while we leave $f_\text{ov}=0.01$ for the advanced stage. A $20$ M$_\odot$ star does not reach the optically thick wind regime for $\Omega/\Omega_c=0.55$ and $f_\text{ov}=0.03$. In contrast, by slightly increasing the overshooting parameter to $f_\text{ov}=0.045$, the same star enters the thick wind regime and self-strips its envelope. The same occurs at $\Omega/\Omega_c=0.5$ and $f_\text{ov}=0.07$. 

The right-hand panel shows that for very large overshooting values $f_\text{ov}=0.109$, even stars with moderate initial rotation $\Omega/\Omega_c=0.45$ can activate optically thick winds. Variations of the overshooting parameter help in reproducing observations of single WRs with lower initial rotations. An accurate investigation of the effect of overshooting and its relation with rotation and metallicity should be carried out with a population study and is deferred to a future work.

\subsection{Thick winds and the HD limit}\label{sec:HD_limit}
We have shown that a few rapidly rotating stars can evolve into single WRs at the metallicity of the SMC as an effect of optically thick winds. We now discuss whether the interplay of enhanced stellar winds and rotation may possibly address the problem of over-luminous cool supergiant stars crossing the HD limit. Answering this question goes beyond the goal of this work and requires a population study, which we will address in a follow-up paper. However, here we briefly speculate how optically thick winds might address the HD limit problem as well. 

\begin{figure*}
    \centering
    \includegraphics[width=0.85\linewidth]{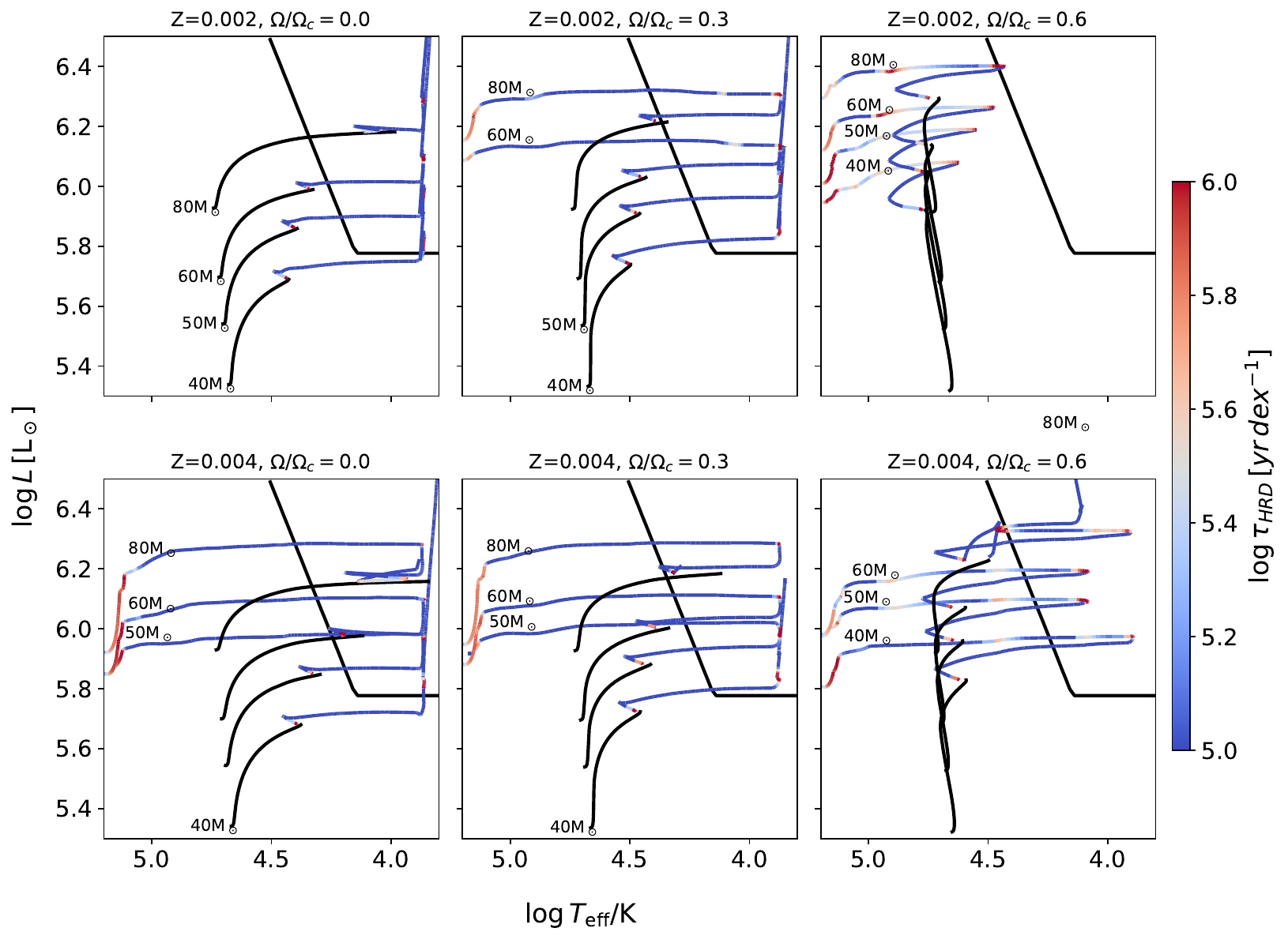}
    \caption{Stellar tracks for stars with mass $M=40,\,50$,\,and $60\,$ M$_\odot$ at $Z=0.002$ (upper panels) and $Z=0.004$ (lower panels), encompassing the metallicity range of the SMC. Columns represents different initial rotations: $\Omega=0$ (left), $\Omega/\Omega_c=0.3$ (middle), $\Omega/\Omega_c=0.6$ (right), corresponding to the minimum, average, and maximum observed rotation velocities of O-type stars in the SMC. The color code shows the time spent in different regions of the HR diagram $\tau_{\rm HRD}$.}
    \label{fig:HD_limit}
\end{figure*}

Figure \ref{fig:HD_limit} shows stellar tracks with different initial rotation velocities and with metallicity $Z=0.002$ and $Z=0.004$ to encompass the metallicity spread in the SMC. Since stars with initial mass $\leq 30$ M$_\odot$ barely enter the HD limit, we focus on higher initial masses ($M=40,$ 50, 60\,\rm M$_\odot$). 

For high initial rotation velocity $\Omega/\Omega_c=0.6$, stars at the lowest considered metallicity ($Z=0.002$) barely enter the HD limit, because of the activation of optically thick winds through channel 1. At higher metallicity, the situation is slightly worse. Indeed, at $Z=0.004$, stars with mass $M\geq 40\,\rm M_\odot$ are able to self-strip their envelope and become WRs. However, since they activate optically thick winds through channel 2, all of them enter the HD limit and spend there a fraction $\sim 2-2.5\%$ of their total lifetime, before moving to the hot part of the HR diagram, where they spend the remaining $\sim 4\%$ of their lifetime.

Reducing the initial rotation worsens the situation significantly for $Z=0.002$ and only slightly for $Z=0.004$. At $\Omega/\Omega_c=0.3$, representing the rotation speed of most O-type stars in the SMC, only stars with $M\geq 60$ M$_\odot$ exit the HD limit at $Z=0.002$. This means that all stars with initial mass $40<M/{\rm M}_\odot<60$ enter and die inside the HD limit. This threshold is lower ($M\geq 50$ M$_\odot$) for $Z=0.004$, but the amount of time spent inside the HD region increases to $\sim 3-6\%$ of the total lifetime, which becomes comparable to the time spent as WR. In the case with no rotation, all the stars with $M\leq 80\,\rm M_\odot$ enter and die inside the HD limit at $Z=0.002$, while the threshold stays at $M\geq 50$ M$_\odot$ for $Z=0.004$. 

Overall, the situation is complex: For high rotation velocities, the HD limit is almost empty at low metallicity ($Z=0.002$), whereas for low or no rotation the HD limit is less populated at high than low metallicity. In any case, stars with initial mass $40<M/{\rm M}_\odot<60$ tend to enter and remain in the HD limit if they are not rotating fast enough. As we have shown in Section~\ref{sec:overshooting}, the situation could be improved by increasing the overshooting parameter, which facilitates activation of optically thick winds through channel 1. A population study such as the one performed by \citet{Gilkis2021}, but with the inclusion of optically thick winds, is required to analyze the situation in more detail, changing metallicity and rotation on a finer grid and testing the effect of different overshooting parameters.

\subsection{Consequences for black hole masses}\label{sec:BH}
\begin{figure}
    \centering
    \includegraphics[width=0.85\linewidth]{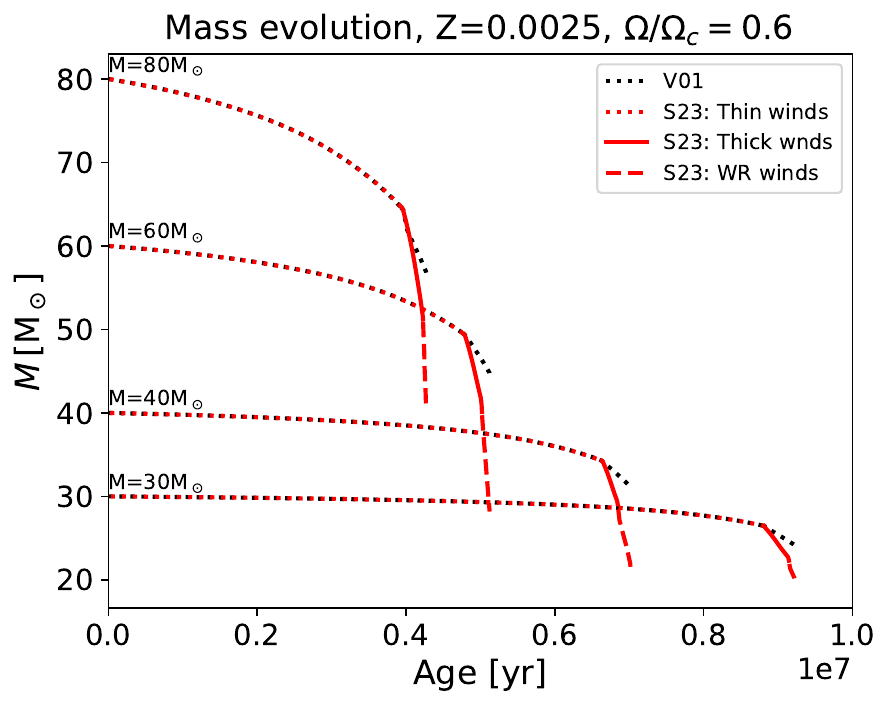}
    \includegraphics[width=0.85\linewidth]{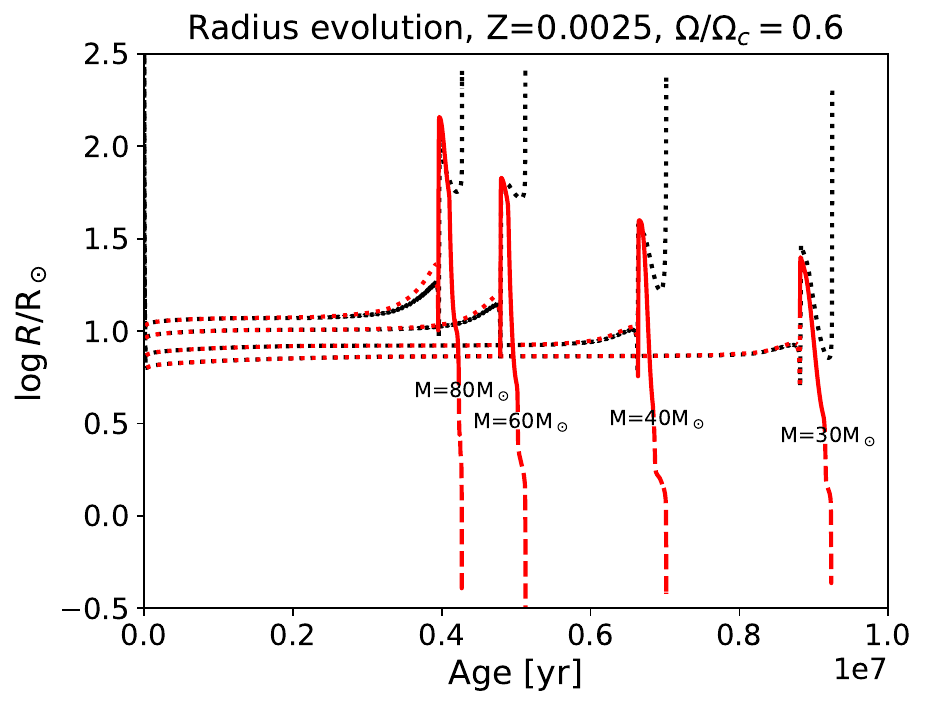}
    \includegraphics[width=0.85\linewidth]{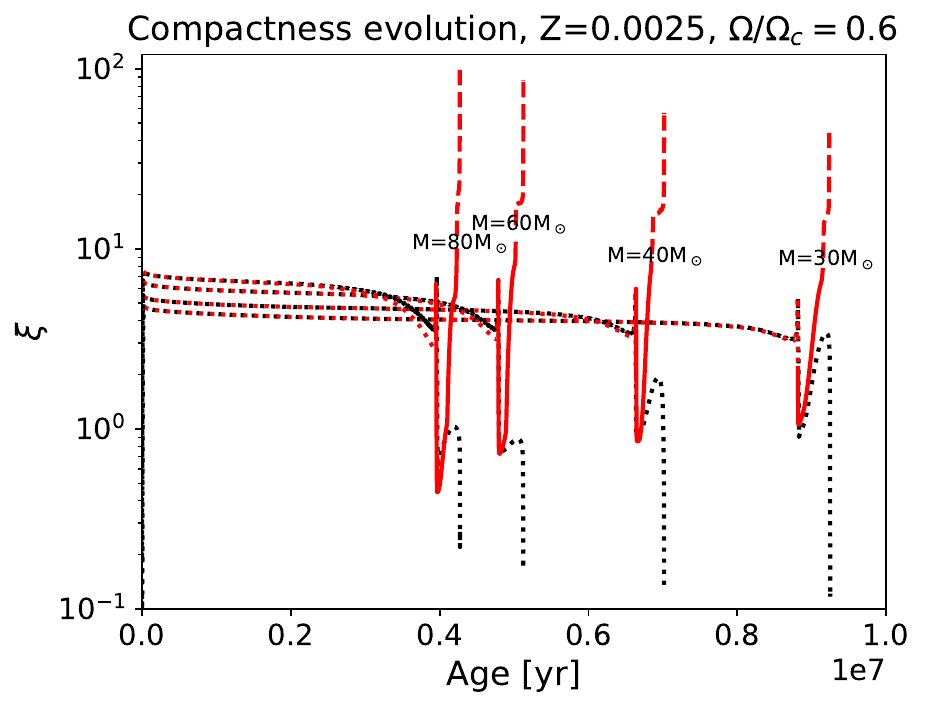}
    \caption{Evolution of stellar mass (top), radius (middle), and envelope compactness (bottom) as a function of the stellar age for $Z=0.0025$ and $\Omega/\Omega_c=0.6$. Black dotted lines: \citetalias{Vink2001} case. Red lines: \citetalias{Sabhahit2023} case. Dotted lines are for optically thin winds, solid lines for optically thick winds, and dashed lines for WR-type winds.}
    \label{fig:massradius_evolution}
\end{figure}

The onset of optically thick winds has severe consequences also for black hole masses, since optically thick winds drastically change the evolution of the stellar mass, radius, and, consequently, of the envelope compactness parameter defined as $\xi\equiv M/R$ \citep[total mass over radius, ][]{fernandez2018}. Figure \ref{fig:massradius_evolution} displays the evolution of these three quantities as a function of time at $Z=0.0025$ and $\Omega/\Omega_c=0.6$. We can see that after the activation of optically thick winds both the stellar mass and radius drop drastically compared to the optically thin wind case. At the end of the evolution, the stellar mass (radius) can be a factor of $\sim 2$ ($\sim 10^3$) smaller in the \citetalias{Sabhahit2023} case compared to the \citetalias{Vink2001} case. This huge variation of the radius is reflected in the compactness parameter, which is extremely high ($\sim 100$) in the \citetalias{Sabhahit2023} case.  
In other words, the WR stars we obtain in the \citetalias{Sabhahit2023} scenario are much more compact (i.e., have a larger value of $\xi{}=M/R$, \citealt{fernandez2018}) compared to the giant stars we obtain as a result of the optically thin wind model.

If the star directly collapses to a black hole at the end of its life, the large  final compactness of our WRs implies that the  mass of the black hole will be very similar to the total mass of the star, because neutrino emission during a failed supernova cannot unbind such a compact envelope \citep{fernandez2018,Costa2022}. In contrast, the low final compactness of the models with optically thin winds implies that a (large) fraction of the envelope might get unbound even in the case of a failed supernova explosion, because of the shock triggered by neutrino emission and propagating through the loosely bound envelope.

\begin{figure}
    \centering
\includegraphics[width=0.85\linewidth]{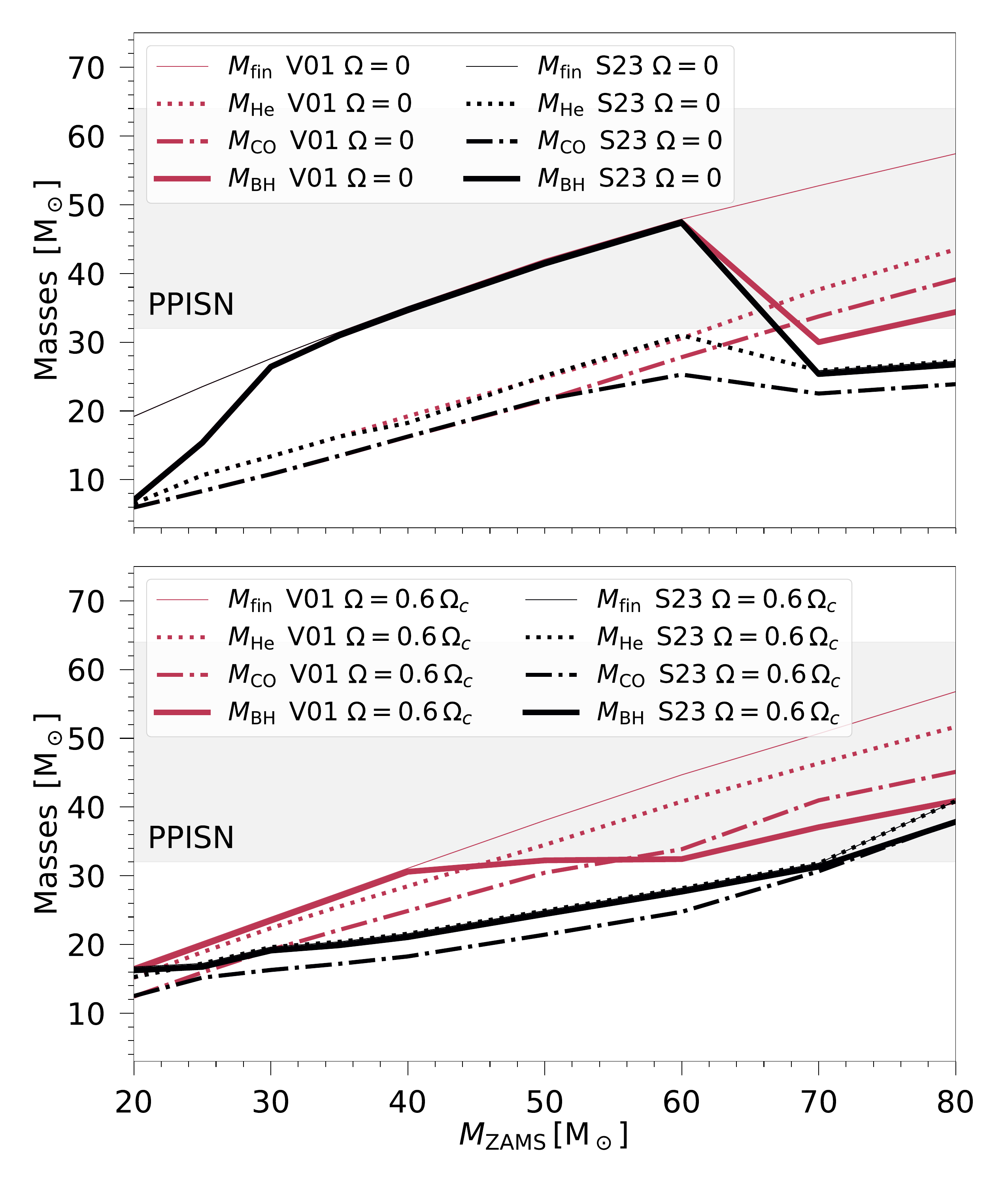}
    \caption{Relation between $M_{\rm ZAMS}$ and final mass of the star ($M_{\rm fin}$, thin solid line), He core mass ($M_{\rm He}$, thick dotted line), CO core mass ($M_{\rm CO}$, thick dot-dashed line), and black hole mass ($M_{\rm BH}$, thick solid line), for optically thin winds (\citetalias{Vink2001}, magenta) and for the \citetalias{Sabhahit2023} model (\citetalias{Sabhahit2023}, black). Upper panel: $\Omega/\Omega_c=0$; lower panel: $\Omega/\Omega_c=0.6$. Gray shaded area: range of $M_{\rm He}$ in which we expect a pulsational pair instability supernova.}
    \label{fig:BH_masses}
\end{figure}

The main uncertainty is whether (and which of) our WRs  collapse directly into black holes or undergo a successful core-collapse supernova. Here, we estimate the explodability of our stellar models with the delayed explosion formalism by \citet{Fryer2012}. According to this simplified model, the final mass of the compact object only depends on the final CO-core mass and total mass of the progenitor star (Figure \ref{fig:BH_masses}). We will perform a more detailed explodability study in a follow-up paper.

Figure~\ref{fig:BH_masses} shows a difference between fast rotating and non-rotating models. For non-rotating models (upper panels), the final stellar mass ($M_{\rm fin}$), final He core mass ($M_{\rm He}$), and final CO core mass ($M_{\rm CO}$) of model \citetalias{Sabhahit2023} are identical to those of model \citetalias{Vink2001} until $M_{\rm ZAMS}<60$ M$_\odot$. In fact, for the metallicity shown in the figure ($Z=0.0025$), non-rotating stars enter the optically thick wind regime only if $M_{\rm ZAMS}\geq{}60$ M$_\odot$. As a consequence, the black hole mass $M_{\rm BH}$ is identical in model \citetalias{Sabhahit2023} compared to \citetalias{Vink2001} up to $M_{\rm ZAMS}=60$ M$_\odot$. The differences in the black hole mass are minimal even for larger ZAMS masses, because pulsational pair instability kicks in at $M_{\rm He}\geq{}32$~M$_\odot$ for the \citetalias{Vink2001} model, removing the residual envelope mass for both model \citetalias{Vink2001} and \citetalias{Sabhahit2023}. The \citetalias{Sabhahit2023} model never enters pulsational pair instability \citep{Simonato2025, Torniamenti2025, Shepherd2025}, but it has already lost all of the H-rich envelope mass.

In the fast rotating case ($\Omega=0.6\,{}\Omega_c$, lower panel of Figure \ref{fig:BH_masses}), the differences in the final masses ($M_{\rm fin}$, $M_{\rm He}$, $M_{\rm CO}$) between models \citetalias{Vink2001} and \citetalias{Sabhahit2023} are large already for $M_{\rm ZAMS}=25$~M$_\odot$. In fact, if the star rotates fast, the optically thick winds activate already at $M_{\rm ZAMS}=25$~M$_\odot$, peeling-off the H-rich envelope. Indeed, the lower panel of Figure \ref{fig:BH_masses} shows that the  final total mass ($M_{\rm fin}$) of a fast rotating star in the \citetalias{Sabhahit2023} model  is almost always the same as its  final He core mass ($M_{\rm He}$). In the rotating case, both models \citetalias{Vink2001} and \citetalias{Sabhahit2023} tend to develop more massive He and CO cores because of rotational mixing. In model \citetalias{Sabhahit2023}, the growth of the He core is limited by core erosion from stellar winds. This explains why the black hole mass of model \citetalias{Vink2001} is larger than that of model \citetalias{Sabhahit2023} for the entire considered mass range. Pulsational pair instability  kicks in at $M_{\rm ZAMS}\geq{}40$ M$_\odot$ and  $M_{\rm ZAMS}\geq{}70$ M$_\odot$ for models \citetalias{Vink2001} and \citetalias{Sabhahit2023}, respectively\footnote{It is important to mention, however, that our model for pulsational pair instability was calibrated for non-rotating stars and must be taken with a grain of salt for rotating models \citep{Spera2017,Mapelli2020b}.}.

The mass and radius evolution of our new models potentially has a large impact on the formation of binary compact objects. On the one hand, stronger winds lead to less bound binaries. This is due to (1) the rapid drop in stellar mass, which increases the binary separation, and (2) the radius evolution, which prevents the star from filling the Roche lobe and reduces the probability of a common envelope. On the other hand, a milder stellar expansion may prevent the merger of close stellar binaries at the end of the main sequence phase, increasing the number of close objects that survive to form a tight compact object binary. The balance of these two effects will be studied in more detail in a future work. 

\subsection{Optically thick winds at lower metallicity}
\begin{figure}
    \centering
    \includegraphics[width=0.9\linewidth]{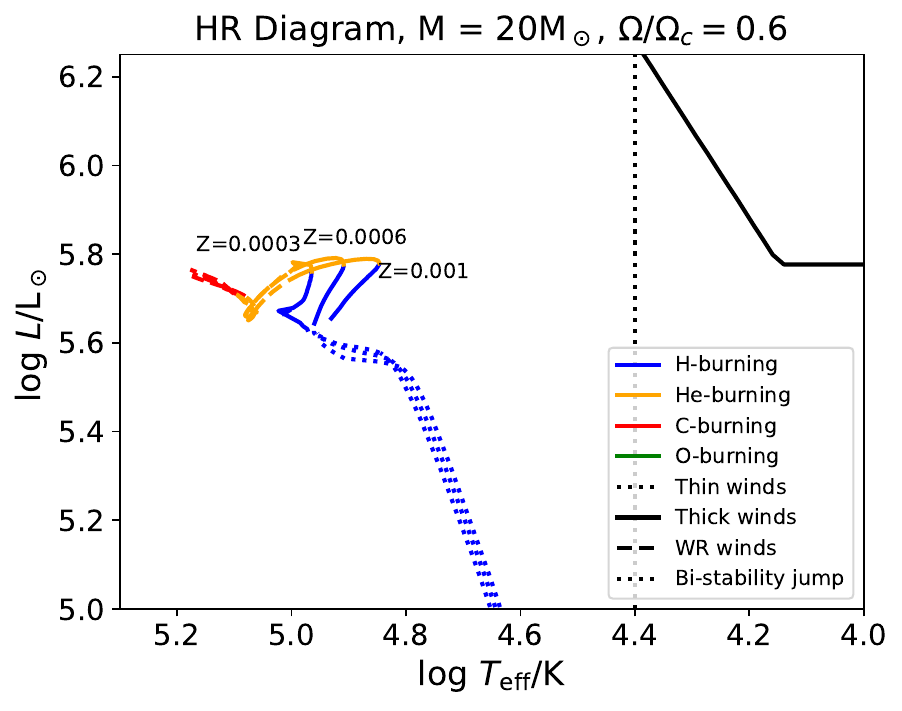}
    \caption{Tracks of a 20 M$_\odot$ star with initial rotation $\Omega/\Omega_c=0.6$ and metallicity $Z=0.001,$ 0.0006, and 0.0003. The line styles and color code are the same as in Figure \ref{fig:new_winds}. We plot the tracks only until the end of C burning; afterward, in the very last stages of their evolution, the $T_{\rm eff}$ and luminosity  oscillate widely in our \textsc{mesa} models, and we did not plot them in order to have a cleaner image (see, e.g., \citealt{Costa2022}). Even at these low metallicities, stars are able to activate thick winds and peel off their envelopes.}
    \label{fig:lowZ}
\end{figure}

To explore whether single WRs could be produced by self-stripping and the activation of the thick wind regime at metallicities even lower than that of the SMC, we simulate the evolution of a $20$ M$_\odot$ star with initial rotation velocity $\Omega/\Omega_\text{c}=0.6$ at $Z=0.001,\,0.0006,\,0.0003$, corresponding roughly to $Z_\odot/15$, $Z_\odot/25$, and $Z_\odot/50$. The latter corresponds to the metallicity of IZw18, for which WR signatures have been observed in the past \citep{Izotov1997} and chemically homogeneous evolution caused by strong initial rotation has been suggested as a possible channel to create these \citep[e.g.,][]{Szecsi2015,Szecsi2025,Kubatova2019}. Figure \ref{fig:lowZ} shows that even at these low metallicities, rotating stars are indeed able to activate the thick wind regime according to the \citetalias{Sabhahit2023} prescription and thus would be able strip their envelope. This happens because stars at lower metallicities evolve on more luminous and compact tracks, facilitating the activation of thick winds through channel 1. Notably, the different wind prescription leads to a distinctively different shape of our tracks compared to early studies simulating chemically homogeneous evolution caused by high initial rotation \citep[e.g.][]{Maeder1987,Yoon2005,Szecsi2015}. The characteristic hook-like feature when leaving the main sequence is not observed in older studies as they tend to assume a generally stronger wind mass-loss, while with the new treatment from \citetalias{Sabhahit2023}, only the activation of optically thick winds keeps the stars eventually compact. Nonetheless,
assuming the high amount of initial rotation assumed in our models is realized in nature, our models confirm a potential route for generating single WRs in extremely metal-poor galaxies such as IZw18 \citep{Legrand1997, Aloisi1999, Schaerer1999, Shirazi2012, Kehrig2013, Szecsi2015,Szecsi2025, Kubatova2019}.

\section{Summary}\label{sec:summary}
Motivated by the recent confirmation that there is a group of companionless WR stars in the SMC \citep{Schootemeijer2024}, we have created a novel set of stellar evolution models to provide a scenario that can explain the self-stripping of these and other stars at low metallicity. This scenario requires the stars to reach a regime of optically thick winds, which we describe with the predictions of \citetalias{Sabhahit2023}, through high initial rotation. Our main results are the following: 
\begin{itemize}
    \item We have shown that optically thick winds (\citetalias{Sabhahit2023}) combined with high initial rotation, $\Omega/\Omega_c\sim 0.6$, are able to strip single metal-poor O-type stars and transform them into WNh stars even at the metallicity of the SMC.
    \item The dependence on metallicity and rotation is not trivial. A metallicity $Z\leq{}0.002$ and high rotation $\Omega{}\gtrsim{}0.6$ $\Omega_c$ enable chemically homogeneous evolution, as the star is more luminous and compact than a slowly rotating star. This facilitates the activation of thick winds by the end of the core H-burning phase (channel 1). In contrast, a higher metallicity ($Z\geq{}0.004$) and lower rotation velocity lead to lower luminosities and effective temperatures and may lead to the activation of thick winds through the bi-stability jump (channel 2). Intermediate metallicity ($Z\sim{0.003}$) and rotation, instead, can lead to the formation of cool supergiant stars.
    \item We find a reasonable agreement between our models and the observed parameters of the hot SMC WNh stars with respect to the time spent in the different regions of the HR diagram, their surface abundance, and rotation velocity. We therefore consider our scenario a plausible path for the formation of single WRs at a lower metallicity.
    \item Within our model setup for the stripping of the stars, overshooting has a similar effect as rotation. For higher values of the overshooting parameter, $f_\text{ov}$, lower initial rotations are required to self-strip the envelope.
    \item The activation of optically thick winds combined with rotation also can help in alleviating the problem of overluminous giants inside the HD limit.
    \item Within our framework, the final stellar masses, core masses, and radii are smaller if the optically thick winds are enabled. This can impact the supernova explosion scenario, the final black hole masses, and the formation of binary compact objects.
    \item Fast rotating massive O-type stars develop thick winds  even at a lower metallicity than the SMC (down to $Z=0.0003\sim{}Z_\odot/50$) and can self-strip their envelope at least until carbon burning. This provides a possible explanation for the presence of WRs in extremely metal poor galaxies as IZw18.
\end{itemize}

\begin{acknowledgements}
We thank the anonymous referee for helpful and constructive comments. We thank Filippo Simonato, Marco Dall'Amico, and Sandro Bressan for useful discussions. MM acknowledges financial support from the European Research Council for the ERC Consolidator grant DEMOBLACK, under contract no. 770017. LB, MM and ST acknowledge financial support from the German Excellence Strategy via the Heidelberg Cluster of Excellence (EXC 2181 - 390900948) STRUCTURES. AACS and VR acknowledge support by the Deutsche Forschungsgemeinschaft (DFG, German Research Foundation) in the form of an Emmy Noether Research Group -- Project-ID 445674056 (SA4064/1-1, PI Sander). This project was co-funded by the European Union (Project 101183150 - OCEANS). ST acknowledges financial support from the Alexander von Humboldt Foundation for the Humboldt Research Fellowship. EK and MM acknowledge support from the PRIN grant METE under contract No. 2020KB33TP. EK acknowledges financial support from the Fondazioni Gini for the Gini grant. GNS ad JSV are supported by STFC funding under grant number ST/Y001338/1. We acknowledge support by the state of Baden-W\"urttemberg through bwHPC and the German Research Foundation (DFG) through grants INST 35/1597-1 FUGG and INST 35/1503-1 FUGG.
We made use of the \textsc{mesa} software (\url{https://docs.mesastar.org/en/latest/}) version r12115; \citep{Paxton2011, Paxton2013, Paxton2015, Paxton2018, Paxton2019}.
We made use of the \textsc{mesa} inlists \url{https://github.com/Apophis-1/VMS_Paper1}, and \url{https://github.com/Apophis-1/VMS_Paper2} from \citet{Sabhahit2023}. This research made use of \textsc{NumPy} \citep{Harris20}, \textsc{SciPy} \citep{SciPy2020}, and \textsc{Matplotlib} \citep{Hunter2007}.

\end{acknowledgements}

\bibliographystyle{aa}
\bibliography{references}

\begin{appendix}
\section{Evolution of $\Gamma_\text{e}$ and $\Gamma_\text{e, switch}$}
\begin{figure}
    \centering
    \includegraphics[width=1.\linewidth]{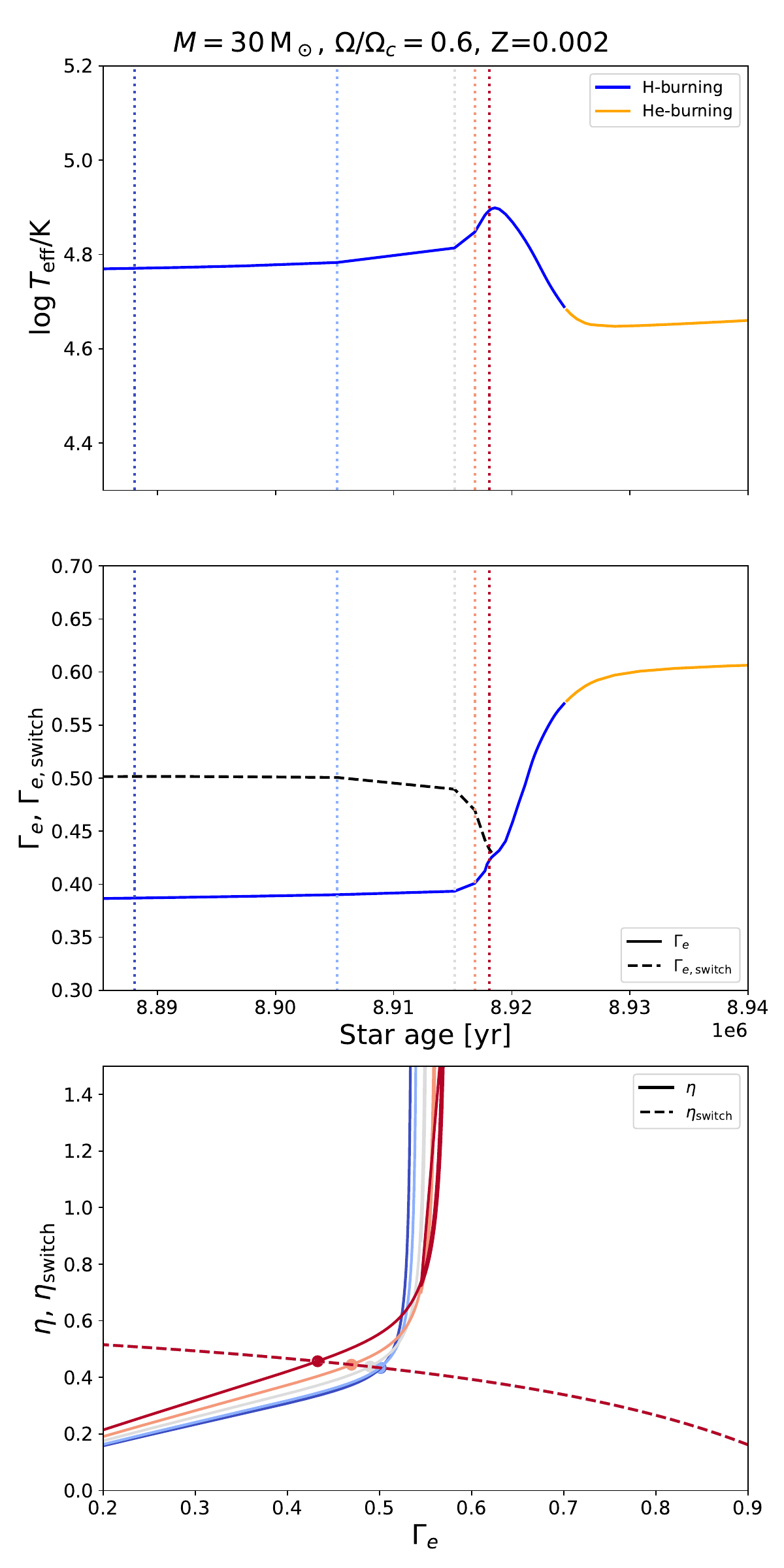}
    \caption{Evolution of $\log (T_{\rm eff}/\textrm{K})$ (top panel), $\Gamma_\text{e}$ and $\Gamma_\text{e, switch}$ (middle panel, solid and dashed lines) as a function of stellar age for a star with $M=30\,\rm M_\odot$, $\Omega/\Omega_c=0.6$, at $Z=0.002$. The color of the solid line indicates the burning stage: blue for hydrogen burning, orange for helium burning. The bottom panel shows $\eta$ and $\eta_{\rm switch}$ as a function of $\Gamma_\text{e}$ at the times indicated by the vertical dotted lines in the top and middle panels. The lines' color goes from blue to red increasing the time. The point where $\eta=\eta_{\rm switch}$ is marked and represents the threshold for thick winds activation $\Gamma_\text{e, switch}$.}
    \label{fig:App_002}
\end{figure}

\begin{figure}
    \centering
    \includegraphics[width=1.\linewidth]{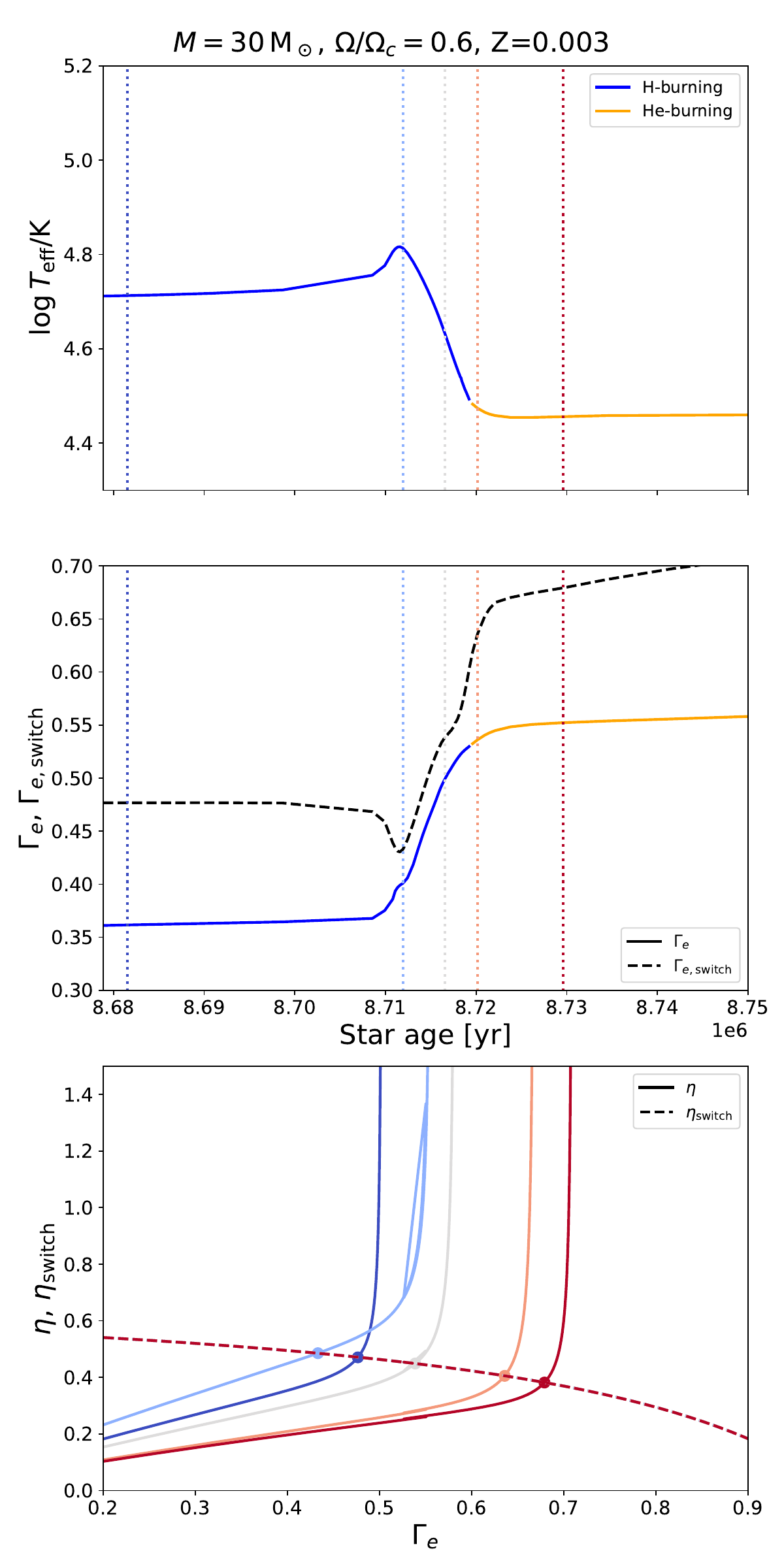}
    \caption{Same as Figure \ref{fig:App_002} but for $Z=0.003$.}
    \label{fig:App_003}
\end{figure}

\begin{figure}
    \centering
    \includegraphics[width=1.\linewidth]{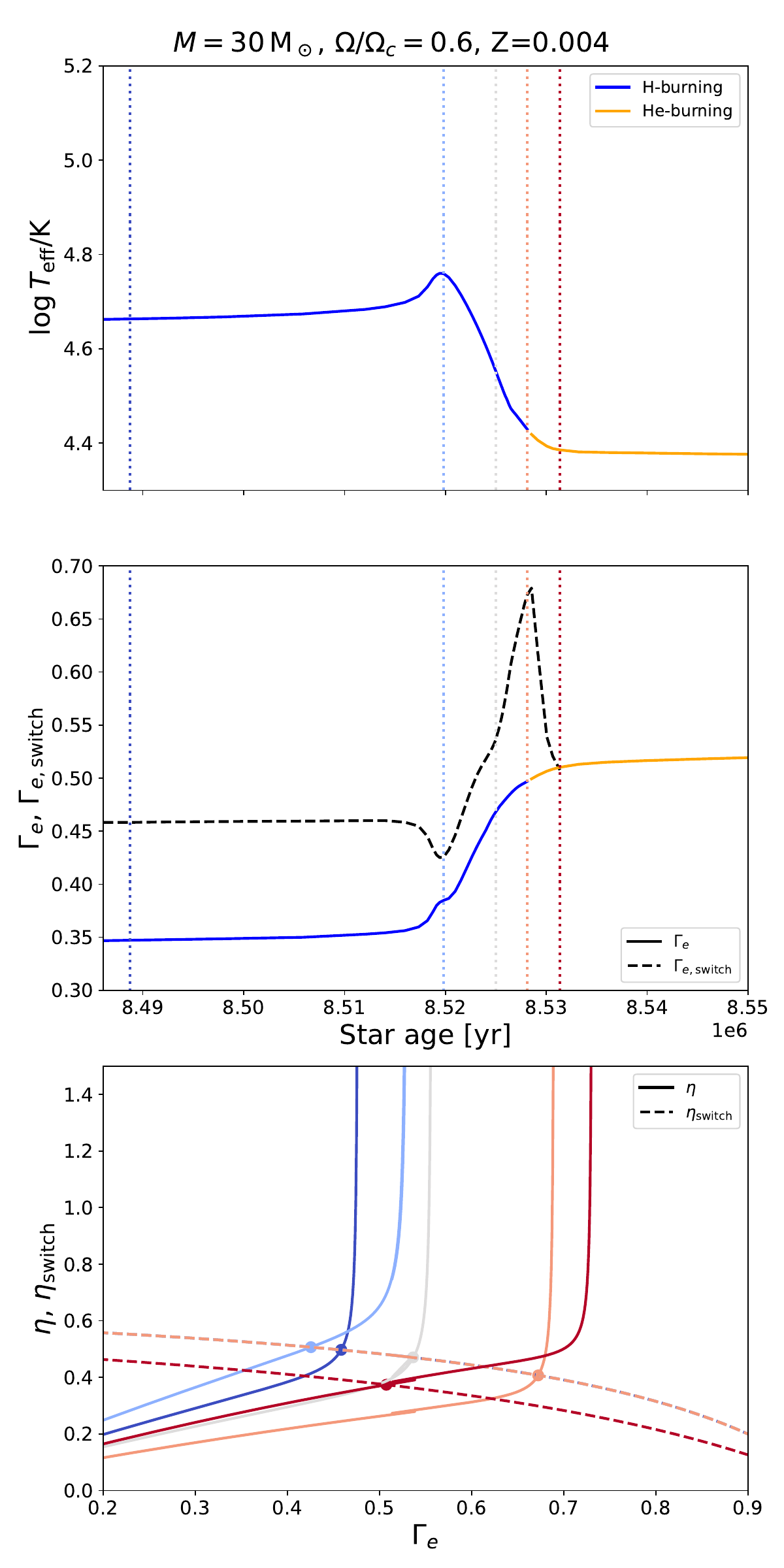}
    \caption{Same as Figure \ref{fig:App_002} but for $Z=0.004$.}
    \label{fig:App_004}
\end{figure}

Here, we show the evolution of $\Gamma_\text{e}$ and $\Gamma_{e, \rm switch}$, as well as of the wind efficiency parameter $\eta$ and $\eta_{\rm switch}$ to better understand the activation of thick winds through different channels. Figures \ref{fig:App_002}, \ref{fig:App_003} and \ref{fig:App_004} are for a star of $30\,\rm M_\odot$ at $Z=0.002,\,0.003,\,0.004$ respectively. 

At $Z=0.002$ (Figure \ref{fig:App_002}), at the end of the hydrogen burning phase, the star contracts, $T_{\rm eff}$ increases, luminosity and hence $\Gamma_\text{e}$ increase, while the threshold for thick winds activation $\Gamma_\text{e, switch}$ decreases. The evolution of $\Gamma_\text{e, switch}$ can be understood from the bottom panel which compares $\eta$ and $\eta_{\rm switch}$ as a function of $\Gamma_\text{e}$ at different times. $\Gamma_\text{e, switch}$ corresponds to the $\Gamma_\text{e}$ where $\eta=\eta_{\rm switch}$. We see that, during the contraction phase at the end of the main sequence, the increase of the wind efficiency parameter $\eta$, due to the growth of $v_\infty$, implies a decrease of $\Gamma_\text{e, switch}$. Optically thick winds are activated during this contraction phase, when $\Gamma_\text{e}>\Gamma_\text{e, switch}$ (middle panel), through channel 1. 

At $Z=0.003$ (Figure \ref{fig:App_003}) the behavior of $\Gamma_\text{e, switch}$ is similar, but $\Gamma_\text{e}$ is lower, due to lower luminosity, and never oversteps $\Gamma_\text{e, switch}$ (middle panel). The contraction is then followed by an expansion, where $\Gamma_\text{e, switch}$ rapidly increases due to the drop of $v_\infty$ and of the rotation boost factor. The activation of helium burning stops the expansion and prevents the star from crossing the bi-stability jump at $\log (T_{\rm eff}/\textrm{K})\simeq 4.4$. In this case optically thick winds are not activated and the star becomes a cool supergiant. 

At $Z=0.004$ (Figure \ref{fig:App_004}) luminosity and $\Gamma_\text{e}$ are even lower and $\Gamma_\text{e}<\Gamma_\text{e, switch}$ during the contraction phase. However, after only $\sim 10^4\,\rm yr$, the star crosses the bi-stability jump, $\Gamma_\text{e, switch}$ drops and optically thick winds are activated. As can be seen from the bottom panel, the effects of the bi-stability jump are: (i) increase of $\eta$ due to an increase in $\dot M$, (ii) decrease of $\eta_{\rm switch}$ due to the increase of the ratio $\varv_\text{esc}^2/\varv_\infty^2$ (see Eq.\,\eqref{eq:eta_s}). The combination of these two effects leads to a drop in $\Gamma_\text{e, switch}$ from $\sim 0.7$ to $\sim 0.5$. Optically thick winds are activated through channel 2 (middle panel). 

\FloatBarrier

\section{Outcomes for different metallicities and rotation}
\begin{figure}[H]
    \centering
    \includegraphics[width=0.75\linewidth]{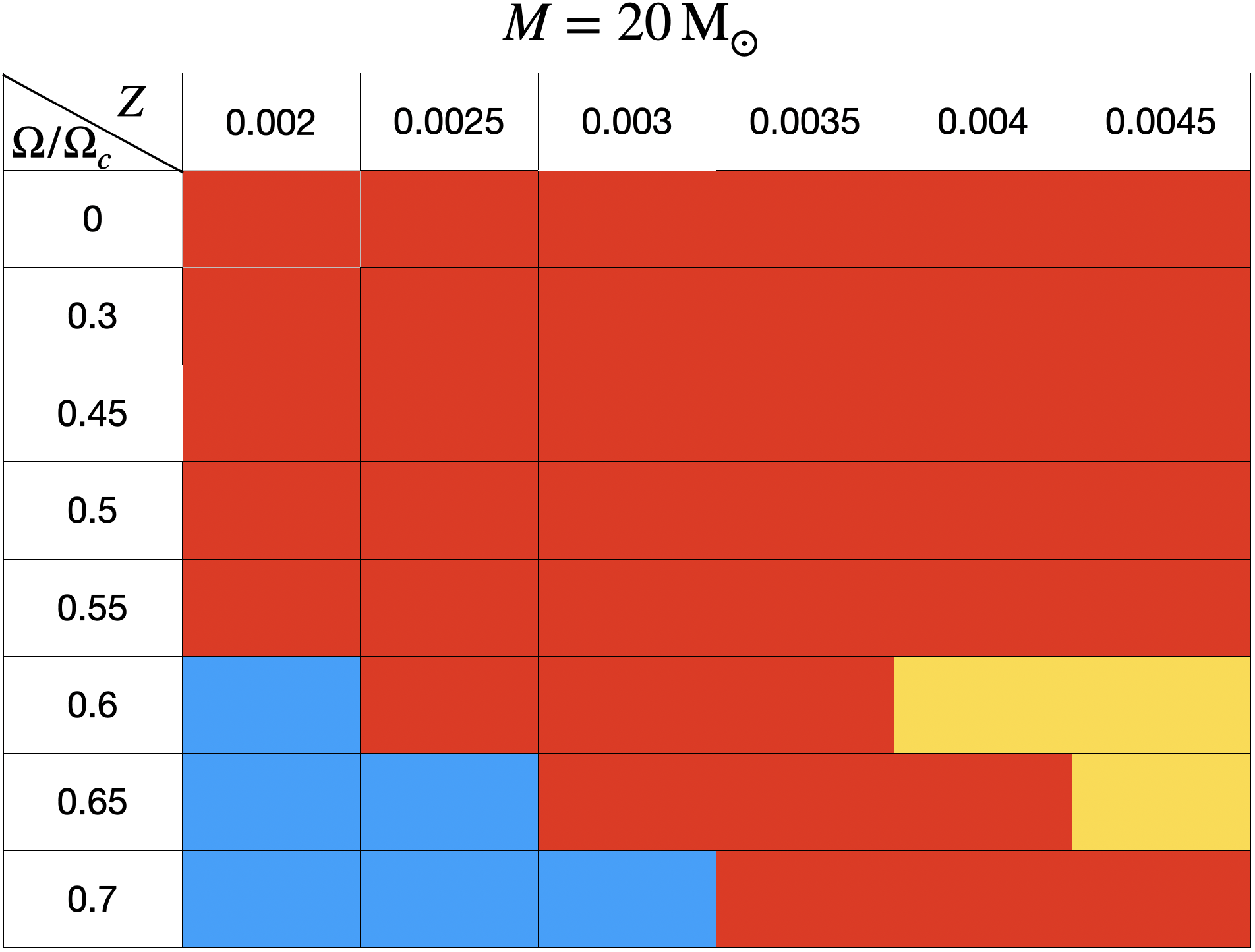}\\[0.35cm]
    \includegraphics[width=0.75\linewidth]{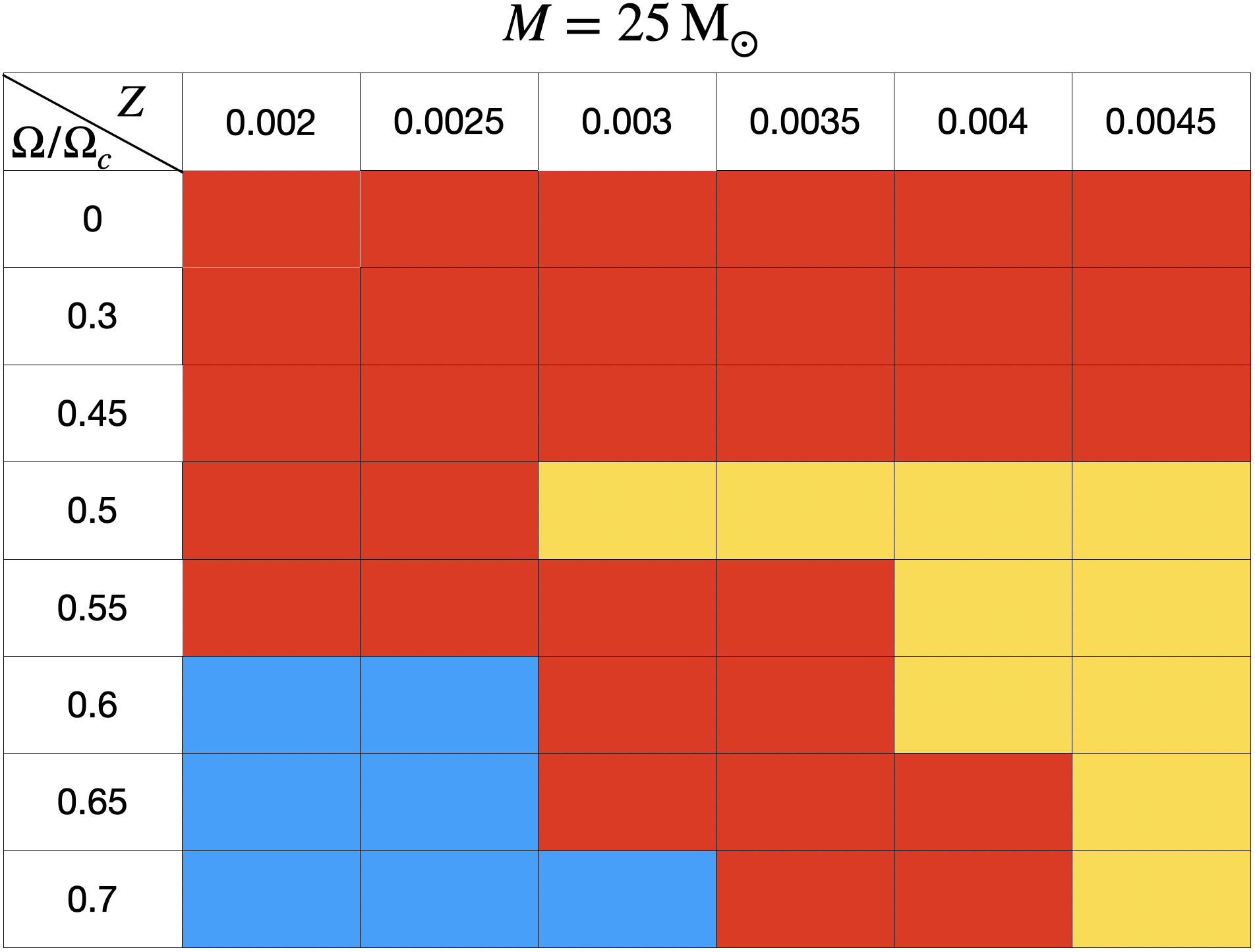}\\[0.35cm]
    \includegraphics[width=0.75\linewidth]{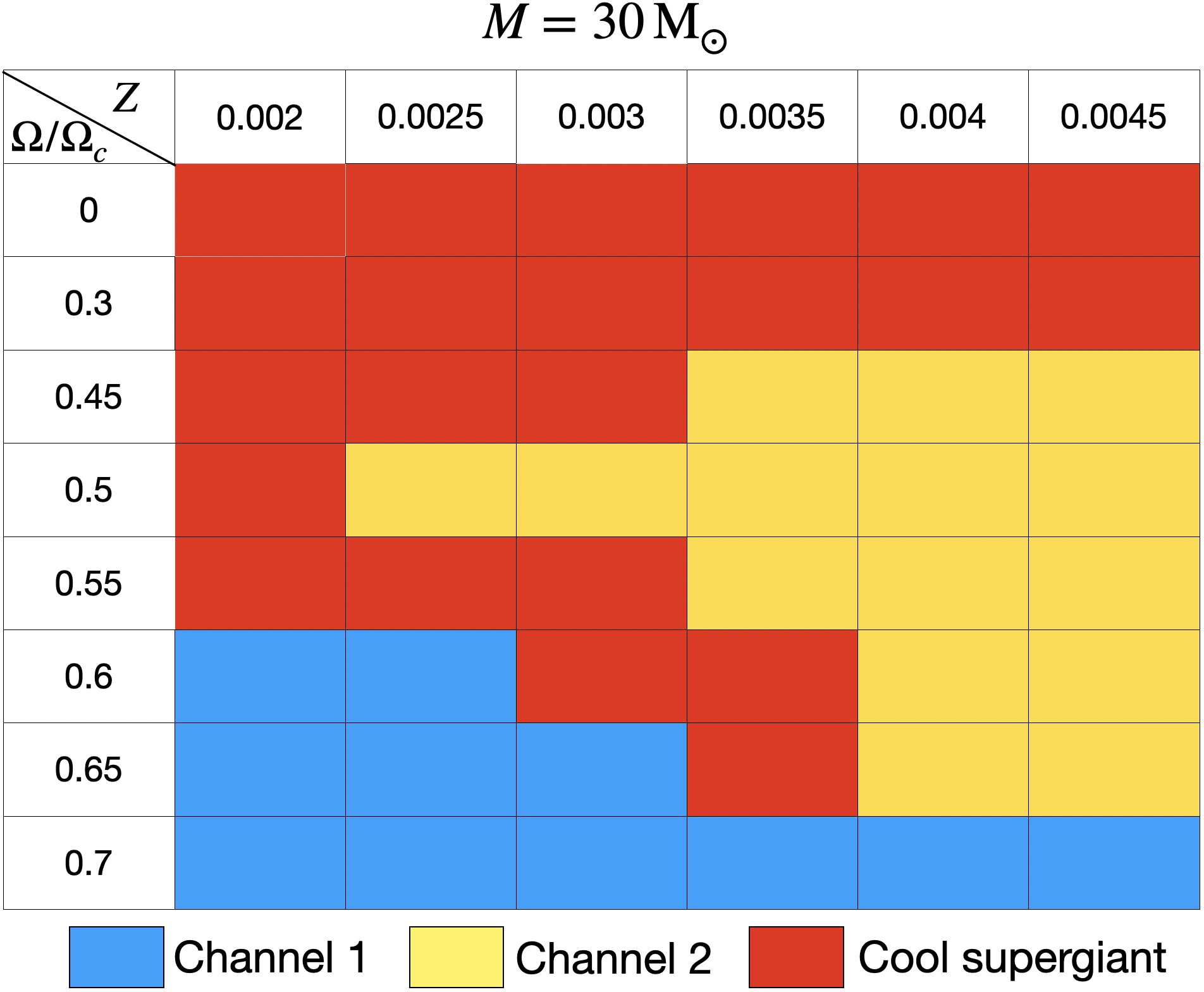}
    \caption{Tables showing the outcome of the simulation for a star with $20\,\rm M_\odot$ (top), $25\,\rm M_\odot$ (middle), $30\,\rm M_\odot$ (bottom), at different metallicities (columns) and initial rotations (rows). The color code represents the outcome: blue for WR formation through channel 1, yellow for WR formation through channel 2, red for cool supergiant formation.}
    \label{fig:tables}
\end{figure}

Figure \ref{fig:tables} displays the outcome of the simulations for different metallicities and rotations for stars with $20,\,25,\,30\,\rm M_\odot$. We see that increasing the initial stellar mass the WR outcome is more and more probable through both channel 1 and 2. For $M=20\,\rm M_\odot$ the minimum rotation required to form a WR is $\Omega/\Omega_c\sim 0.6$, for $M=25\,\rm M_\odot$ this reduces to $\sim 0.5$ and for $M=30\,\rm M_\odot$ to $\sim 0.45$. The dependence on metallicty and rotation is as we described in Section \ref{sec:dependence}. At fixed rotation lower metallicity favors channel 1 activation, high metallicity favors channel 2, and intermediate metallicity leads to a cool supergiant. The same trend can be seen at fixed metallicity and decreasing rotation.

\FloatBarrier

\section{Results for $Z=0.004$, channel 2}\label{app:ch2}
\begin{figure}[H]
    \centering
    \includegraphics[width=0.9\linewidth]{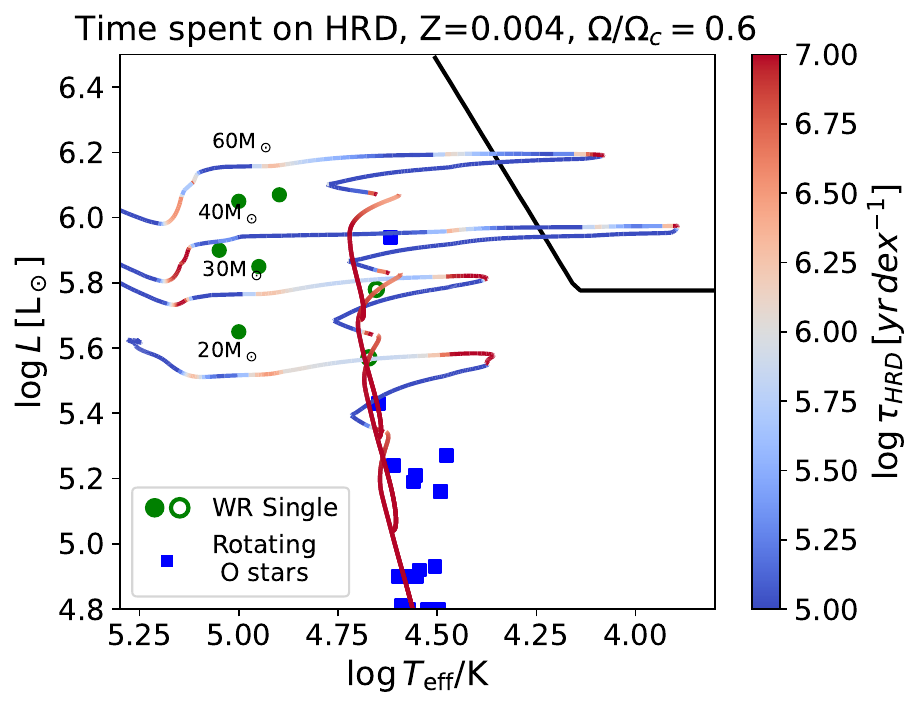}
    \includegraphics[width=0.9\linewidth]{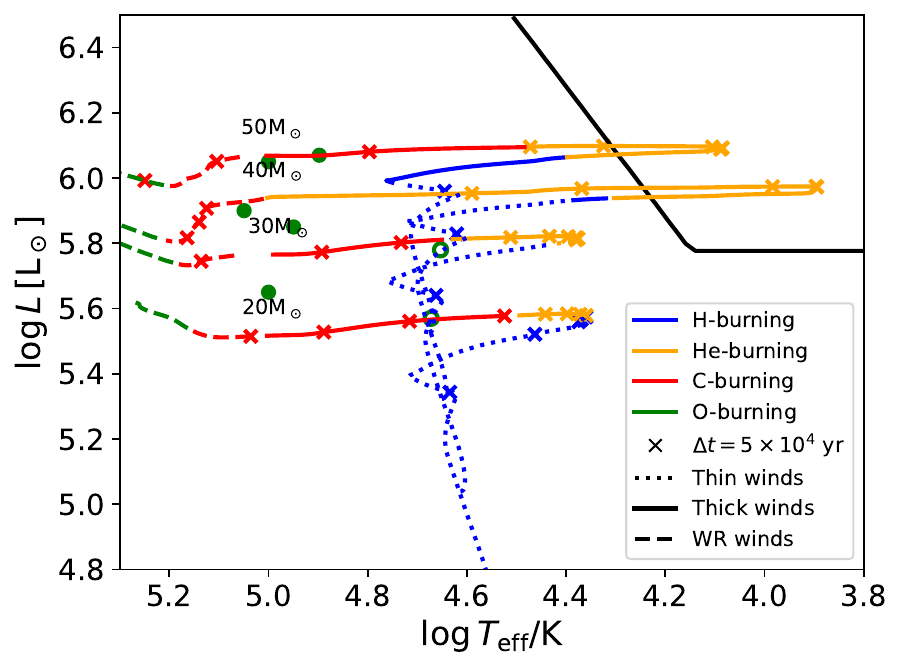}
    \caption{Same as Figure \ref{fig:time} but for $Z=0.004$. Stars in the hot WR phase are generally more evolved with respect to $Z=0.002$ and already in the carbon burning phase.}
    \label{fig:time_004}
\end{figure}

In the main text, we have shown most of the results for $Z=0.002$, where channel 1 activation is favored. In this Appendix, we show the same results for $Z=0.004$, where optically thick winds are activated through channel 2, thanks to the bi-stability jump. Figure \ref{fig:time_004} shows the time spent on the HR diagram for stars with $20,\,30,\,40,\,60\,\rm M_\odot$. As described in Section \ref{sec:dependence}, after hydrogen burning, these stars expand until reaching $\log (T_{\rm eff}/\textrm{K})\sim 4.4$, where they enter the optically thick wind regime thanks to the bi-stability jump. Most of their He-burning phase is spent at lower temperatures $\log (T_{\rm eff}/\textrm{K})<4.6$, and they reach the hot WR stage during core carbon burning.

\begin{figure}[H]
    \centering
    \includegraphics[width=0.9\linewidth]{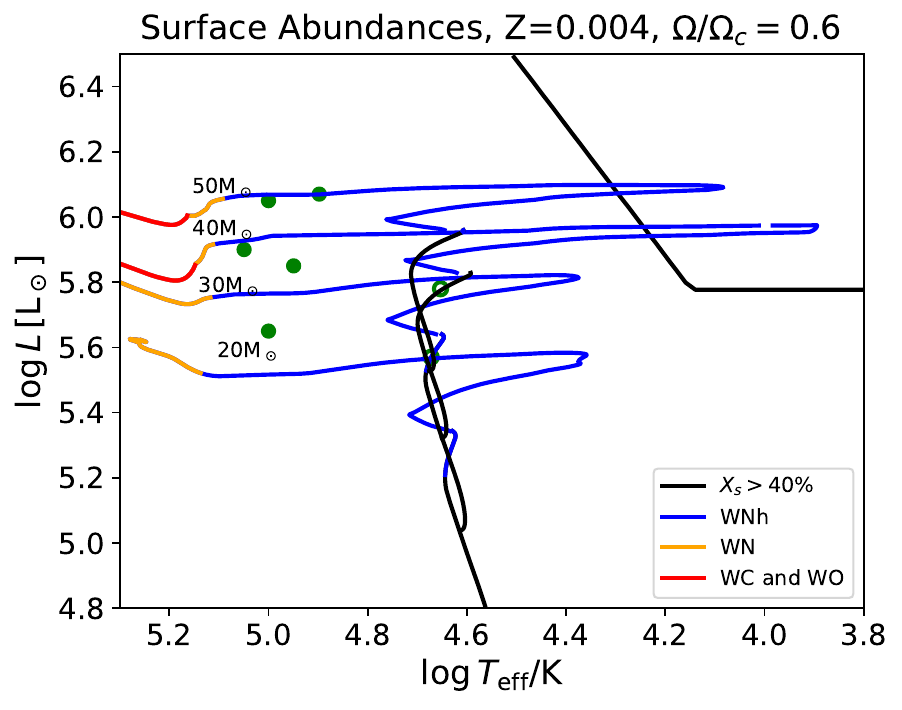}
    \includegraphics[width=0.9\linewidth]{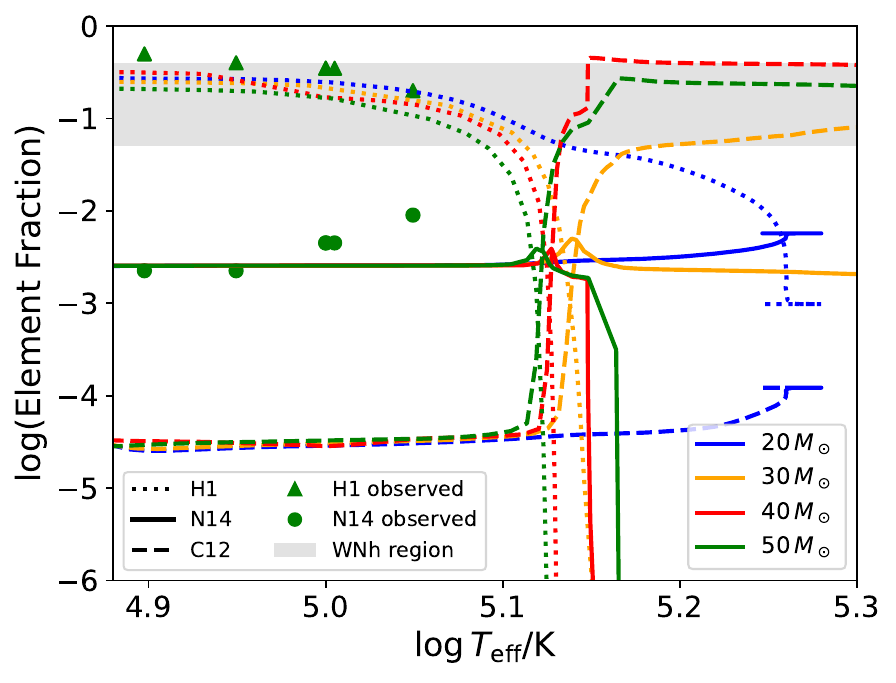}
    \caption{Same as Figure \ref{fig:abundances} but for $Z=0.004$.}
    \label{fig:abundances_004}
\end{figure}

Figure \ref{fig:abundances_004} shows element surface abundances. All the stellar tracks are still compatible with our definition of WNh stars when they pass through the data points. Their are less hydrogen depleted then the $Z=0.002$ case, with surface hydrogen in the range $20-30\%$. Nitrogen is also slightly more abundant, with equilibrium value $\sim 2\times 10^{-3}$, in great agreement with data at $\log (T_{\rm eff}/\textrm{K})<5$. However, even in this case, further nitrogen surface enrichment occurs at higher temperatures with respect to observations.

\begin{figure}
    \centering
    \includegraphics[width=0.9\linewidth]{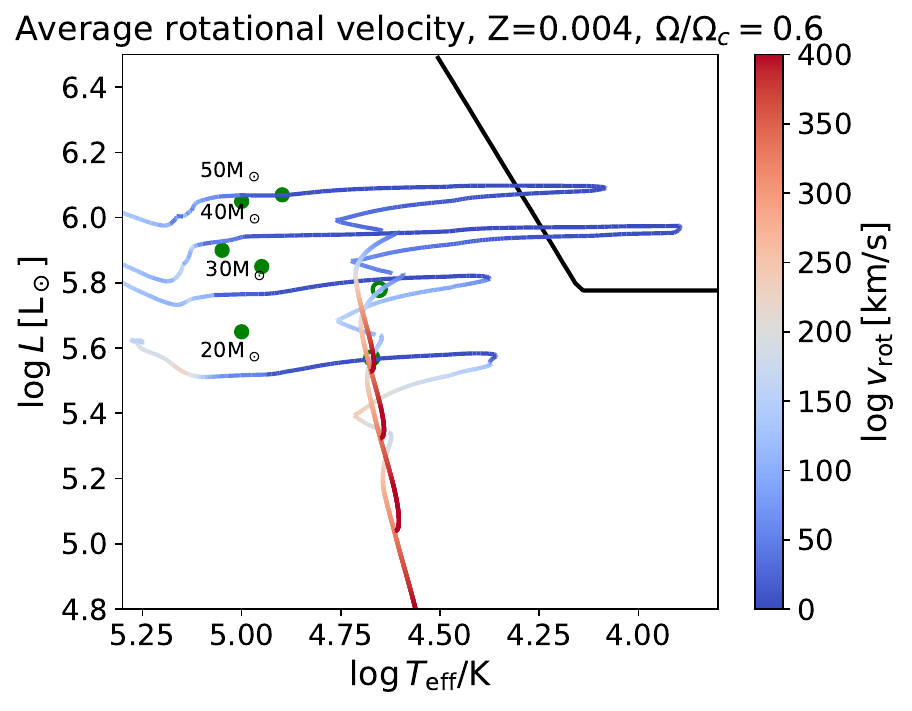}
    \includegraphics[width=0.9\linewidth]{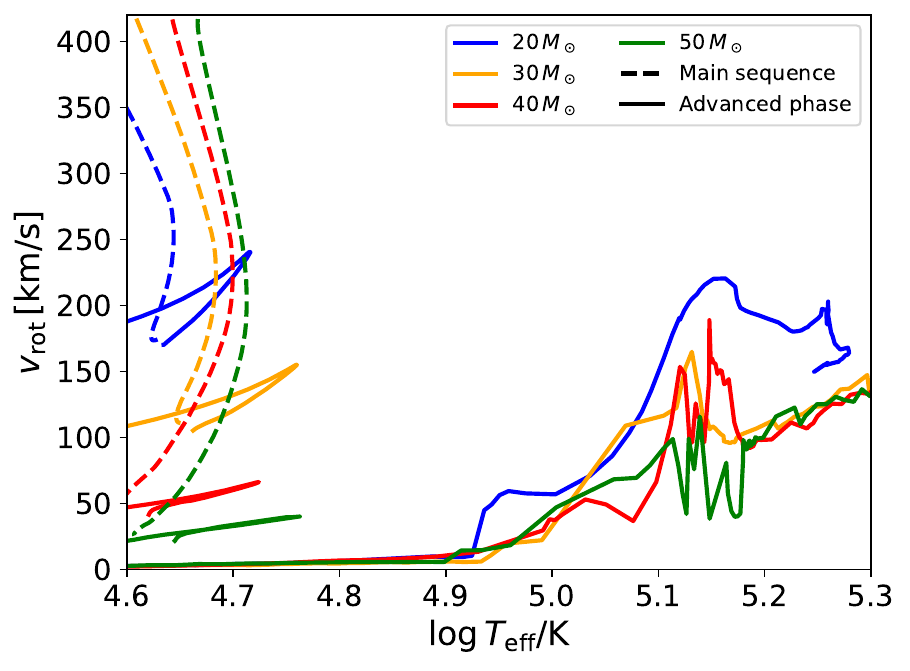}
    \caption{Same as Figure \ref{fig:vsurf_004} but for $Z=0.004$.}
    \label{fig:vsurf_004}
\end{figure}

Stellar rotation is shown in Figure \ref{fig:vsurf_004}. The general behavior is similar to the $Z=0.002$ case, but the drop during the main sequence is more pronounced. While stars expand to lower temperatures, their rotation drops further, and when they transition to the hot WRs phase they are practically non-spinning. Rotation starts to rise again at $\log (T_{\rm eff}/\textrm{K})>4.9$ but in a shallower way with respect to the $Z=0.002$ case. Our models predicts $v_{\rm rot}\sim 50\,\rm km/s$ at $\log (T_{\rm eff}/\textrm{K})\sim 5$, still consistent with the upper limits mentioned in the text.

\begin{figure}
    \centering
    \includegraphics[width=1.\linewidth]{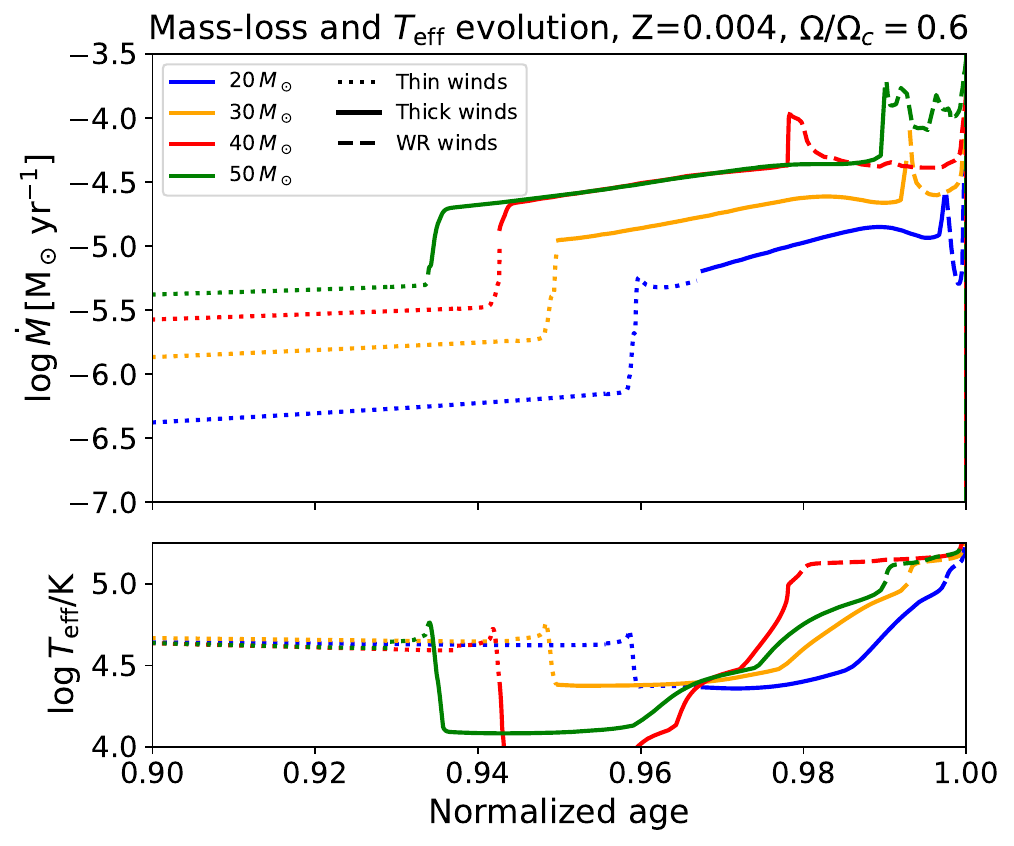}
    \caption{Same as Figure \ref{fig:mass-loss} but for $Z=0.004$.}
    \label{fig:mass-loss_004}
\end{figure}

\begin{figure}
    \centering
    \includegraphics[width=0.9\linewidth]{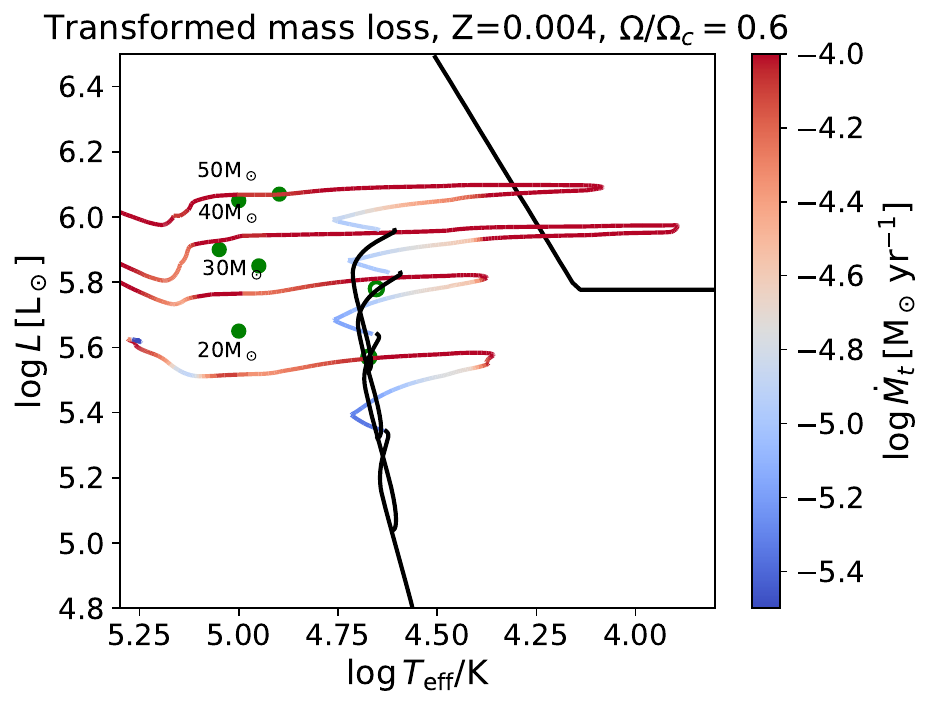}
    \includegraphics[width=0.9\linewidth]{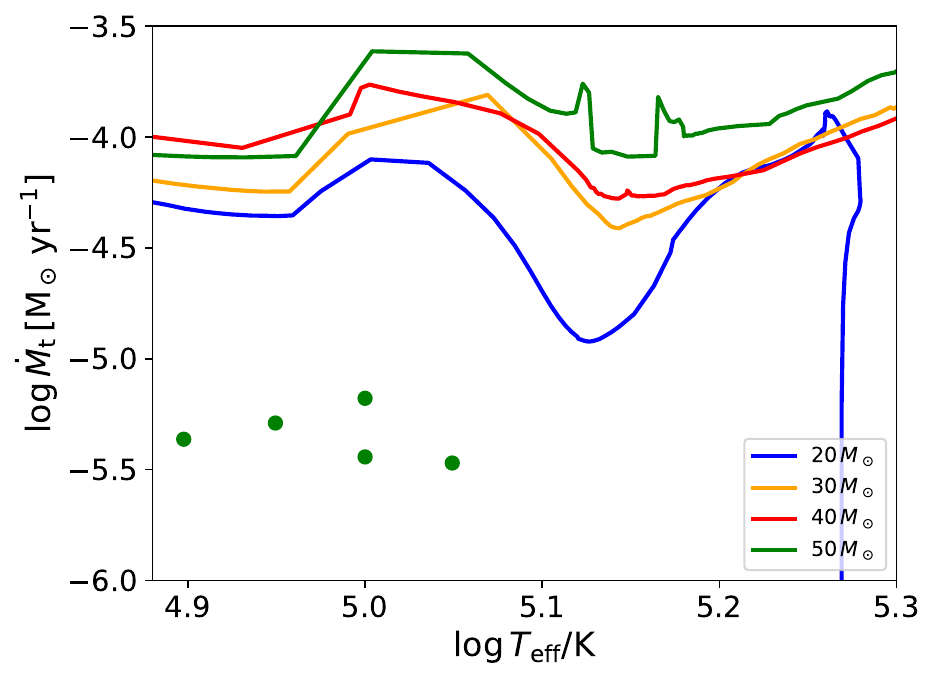}
    \caption{Same as Figure \ref{fig:transformed_massloss} but for $Z=0.004$.}
    \label{fig:transformed_massloss_004}
\end{figure}

The mass-loss and transformed mass-loss are displayed in Figures \ref{fig:mass-loss_004} and \ref{fig:transformed_massloss_004}. Also in this case, the predicted transformed mass-loss is larger than observational data.

\FloatBarrier

\section{Stellar interiors: Abundance profiles}\label{app:profile}
\begin{figure*}
    \centering
    \begin{minipage}[b]{0.495\textwidth}
    \includegraphics[width=1.\linewidth]{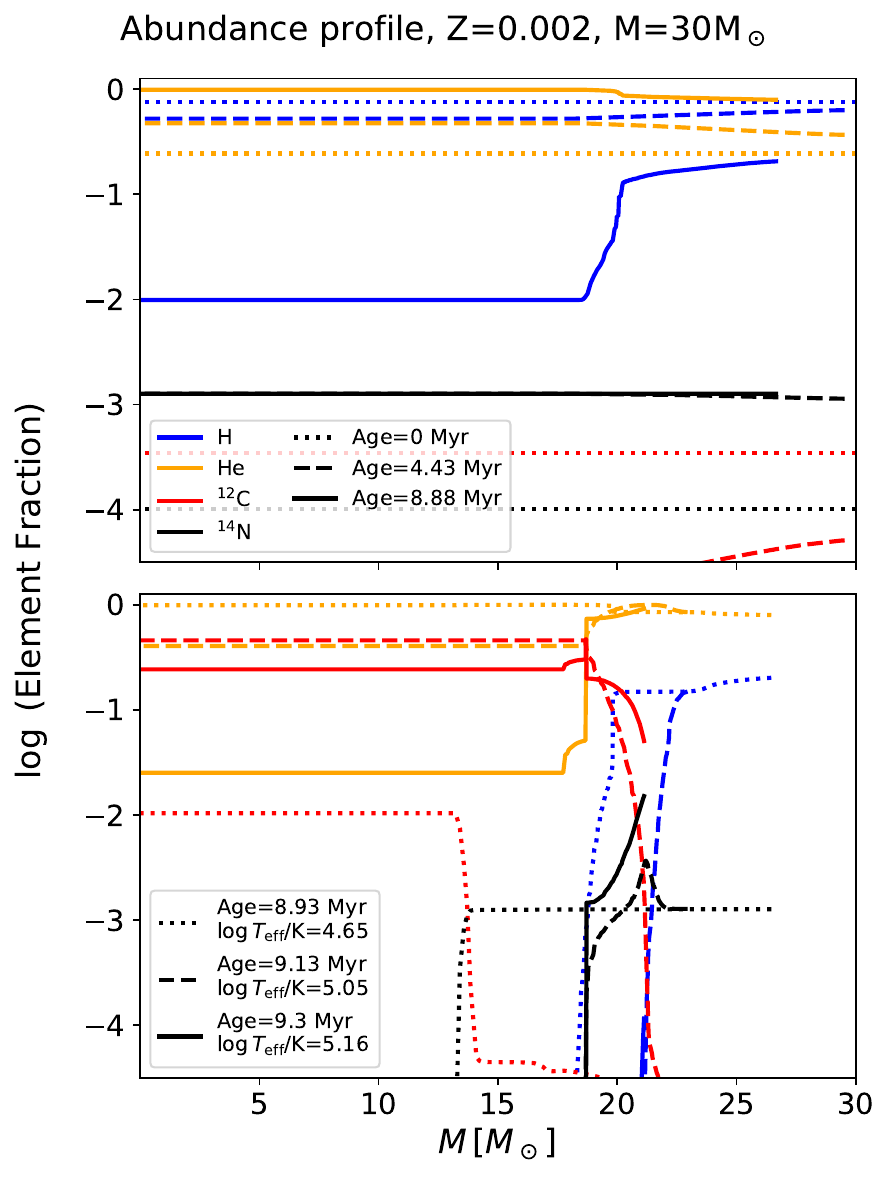}
  \end{minipage}
  \hfill
  \begin{minipage}[b]{0.495\textwidth}
    \includegraphics[width=1.\linewidth]{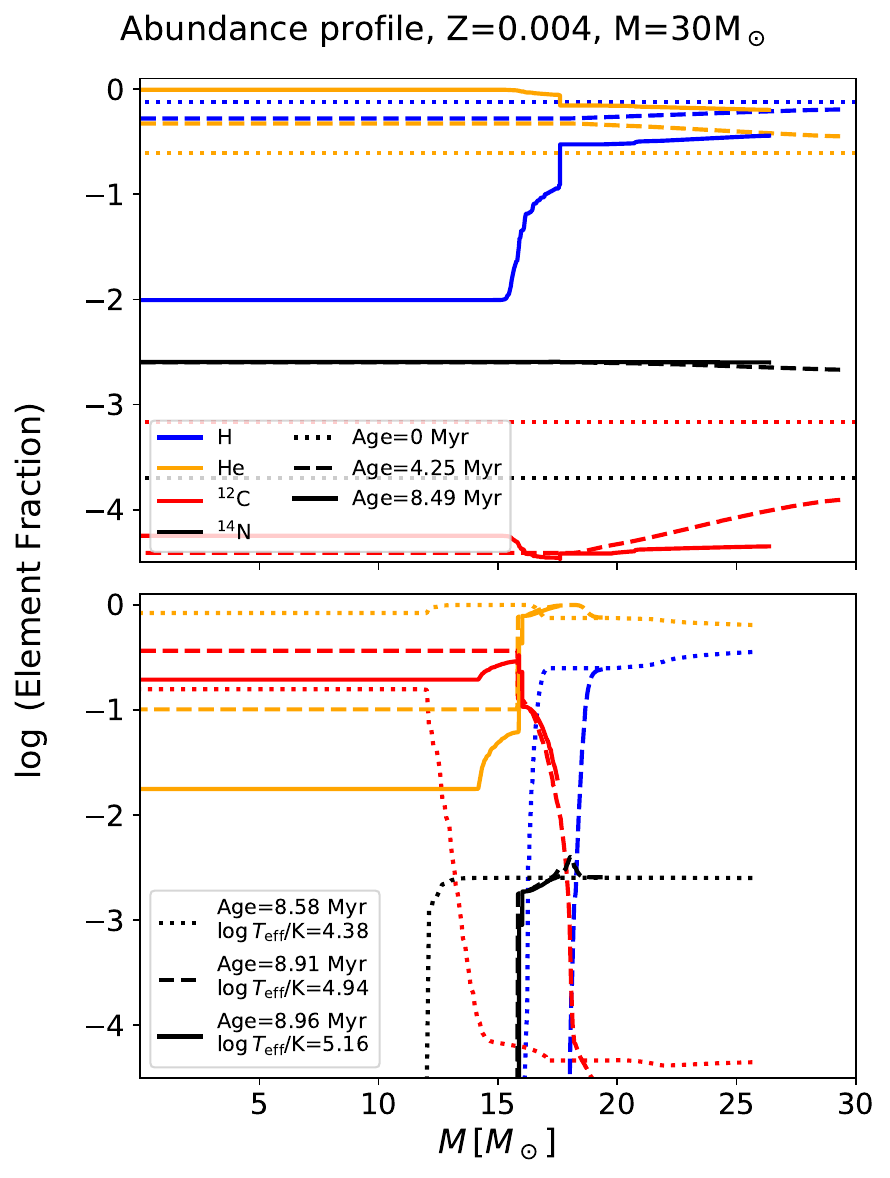}
  \end{minipage}
    \caption{Stellar profiles for a $30\,\rm M_\odot$ star. Abundances of hydrogen (blue), helium (orange), $^{12}$C (red), and $^{14}$N (black) as a function of the mass enclosed in a given shell. Left panels are for $Z=0.002$, right for $Z=0.004$. Top panels: main sequence evolution. Line styles are for different times: dotted for ZAMS, dashed for half main sequence, solid for end of the main sequence. Bottom panels: post main sequence phases. Line styles are for different times: dotted for the time corresponding to minimum temperature, dashed for $\log (T_{\rm eff}/\textrm{K})\sim 5$, solid for $\log (T_{\rm eff}/\textrm{K})\sim 5.15$.}
    \label{fig:profile}
\end{figure*}

Figure \ref{fig:profile} shows the evolution of the abundance profiles of our simulated stars. We take as an example a star with initial mass $M=30\,\rm M_\odot$, representative of those passing through observational data, for $Z=0.002$ (left) and $Z=0.004$ (right), during main sequence (top) and during more advanced phases (bottom). For the main sequence evolution we plot the initial configuration at ZAMS, the final configuration, when central hydrogen drops below $10^{-2}$ ($8.8\,\rm Myr$ for $Z=0.002$, $8.49\,\rm Myr$ for $Z=0.004$), and the configuration at half main sequence time. 

The evolution during the first half of the main sequence is almost purely homogeneous, with helium and nitrogen increasing and hydrogen and carbon decreasing throughout the whole the star. Nitrogen abundance has already reached equilibrium at $10^{-3}$ and $2\times 10^{-3}$ for the two cases. Afterward, pure homogeneity is broken, due to rotational velocity reduction, and some hydrogen remains on the surface $\sim 20\%$ for $Z=0.002$, and $\sim 30\%$ for $Z=0.004$. The helium core, defined as the region where H abundance $<10^{-2}$ and He abundance $>10^{-2}$, is slightly larger in the low metallicity case, $\sim 18\,\rm M_\odot$ against $\sim 15\,\rm M_\odot$.

As for the post main sequence phase, we show 3 times, corresponding to:
\begin{enumerate}
   \item the time when the effective temperature is at its minimum, i.e., the star is in the coldest region of the HR diagram, soon after the end of hydrogen burning, corresponding to $\log (T_{\rm eff}/\textrm{K})\sim 4.65$ (age $8.93\,\rm Myr$) and $\log (T_{\rm eff}/\textrm{K})\sim 4.38$ (age $8.58\,\rm Myr$);
   \item the time when $\log (T_{\rm eff}/\textrm{K})\sim 5$, i.e., most of the observed SMC WRs are found, corresponding to $9.13\,\rm Myr$ and $8.91\,\rm Myr$;
   \item the time when $\log (T_{\rm eff}/\textrm{K})\sim 5.15$, corresponding to very late evolutionary stages ($9.32\,\rm Myr$ and $8.96\,\rm Myr$).
\end{enumerate}
First, we notice that stellar mass is fastly reduced by winds, with both stars loosing $\sim 7\,\rm M_\odot$ in $\sim 0.4\,\rm Myr$. Second, we see that hydrogen is still present on the surface at $\log (T_{\rm eff}/\textrm{K})\sim 5$, making the star appear as a WNh, but it is rapidly eroded by winds. Third, we notice that core helium burning quickly depletes nitrogen in the core. However, in the outer layers, the CNO equilibrium value of nitrogen remains constant even at $\log (T_{\rm eff}/\textrm{K})\sim 5$, in tension with observational data that show nitrogen enhancement at these temperatures.

Both cases present a small nitrogen bump just beneath the surface, but probably rotation is too low to bring it on the surface. In our models, nitrogen enhancement occurs later, at $\log (T_{\rm eff}/\textrm{K})\sim 5.15$, when winds reveal the inner nitrogen bump.

\FloatBarrier

\section{Discussion on the bi-stability jump}\label{app:jump}
The existence and strength of the mass-loss increase found by \citet{Vink1999} has been an object of investigation over the last 25 years, with very different results based on the adopted stellar atmosphere model. While studies using a version of the \textsc{fastwind} code aiming for a locally consistent solution of the wind dynamics find a steady decline of mass-loss rate with $T_{\rm eff}$ with no bi-stability jump at $T_{\rm eff}\sim 25\,\rm kK$  \citep{Bjorklund2021, Bjorklund2023}, \citet{Petrov2016}, using a globally consistent approach with the \textsc{cmfgen} code, confirm the existence of a jump, albeit at a somewhat lower $T_{\rm eff}\sim 20\,\rm kK$. Using a different code (\textsc{metuje}), \citet{Krticka2021,Krticka2024} do find an increase of the mass-loss at cooler $T_\text{eff}\sim 19\rm kK$, but less pronounced than \citet{Vink1999}.

Observationally, changing wind parameters and specific line profiles in individual luminous blue variable stars provides evidence for a mass-loss rate change \citep{Groh2011a, Groh2011b}. This could also be found in recent hydrodynamical evolution models \citet{Grassitelli+2021} implementing a specific formalism of the bi-stability jump mass-loss description \citet{Smith+2004}. In contrast, empirical evidence for a mass-loss rate increase in normal OB supergiants is much less clear. Already \citet{Vink2000} noted a discrepancy between their theoretical predictions and some of the measurements by \citet{Kudritzki1999} in the $12500<T_{\rm eff}/\rm K<22500$ temperature regime. Also \citet{Trundle2004} confirmed the discrepancy between theoretical mass-loss rates and empirical modeling. Recent investigations of blue supergiants in the Galaxy \citep{Bernini-Peron2023,deBurgos2024}, LMC \citep{Verhamme2024,Alkousa2025}, and SMC \citep{Bernini-Peron2024} could not find a clear mass-loss enhancement when transitioning to cooler $T_{\rm eff}$, but in particular the detailed analysis efforts of individual targets \citep{Bernini-Peron2023,Bernini-Peron2024,Alkousa2025} show that there is  no clear monotonic downward trend in the mass-loss rates predicted by \citet{Bjorklund2023}. Among the available descriptions the results align reasonably well with the predictions from \citet{Krticka2021, Krticka2024} including a shallow mass-loss increase, albeit with some scatter depending on the sample. 

However, in the context of the \citetalias{Sabhahit2023}-formalism, it is not the direct $\dot{M}$-jump in the \citet{Vink2001} description, but actually the reduction in the $\eta_{\rm switch}$-value that is the major driving factor for the activation of the thick wind regime. This reduction is caused by a drop of the $\varv_\infty/\varv_{\rm esc}$ ratio. Despite some uncertainty in the stellar masses, which can affect the determination of $\varv_\text{esc}$, a decline of $\varv_\infty/\varv_{\rm esc}$ is well observed, also across different metallicities \citep[e.g.,][]{Lamers1995, Crowther2006, Markova2008,Bernini-Peron2024,Alkousa2025}. Depending on the sample it is currently unclear whether there is a sharp drop in this ratio or a more gradual decline, but in each case the \citetalias{Sabhahit2023}-condition would be met. While this does not imply that channel 2 must occur in nature, its activation criteria is not in conflict with observations and thus justified within our framework.
\end{appendix}
\end{document}